\documentstyle[psfig]{l-aa}               % LaTeX A&A  Standard Fonts
\def\mkmsyr{\mbox{M$_\odot \mbox{km s}^{-1} \mbox{yr}^{-1}$}}
\def\kms{\mbox{km s$^{-1}$}}

\def\myr{\mbox{M$_\odot \mbox{yr}^{-1}$}}

\def\cm2{\mbox{$\mbox{cm}^{-2}$}}
\def\cm3{\mbox{$\mbox{cm}^{-3}$}} 
\def\sm{\mbox{M$_\odot$}}
\def\sl{\mbox{L$_\odot$}}

\def\lbol{\mbox{$L_{\mbox{\tiny bol}}$}}
\def\lacc{\mbox{$L_{\mbox{\tiny acc}}$}}

\def\l13{\mbox{$L_{\mbox{\tiny 1.3mm}}$}}
\def\k13{\mbox{$\kappa_{\mbox{\tiny 1.3mm}}$}}

\def\fco{\mbox{$F_{\mbox{\tiny CO}}$}}

\def\f800{\mbox{$F_{\mbox{\tiny 800}}$}}
\def\fent{\mbox{$f_{\mbox{\tiny ent}}$}}

\def\vwin{\mbox{$V_{\mbox{\tiny w}}$}}

\def\ms{\mbox{$M_{\mbox{\tiny $\star$}}$}}

\def\menv{\mbox{$M_{\mbox{\tiny env}}$}}

\def\macc{\mbox{$\dot{M}_{\mbox{\tiny acc}}$}}

\def\mwin{\mbox{$\dot{M}_{\mbox{\tiny w}}$}}
\def\mw{\mbox{$\dot{M}_{\mbox{\tiny w}}$}}

\def\>{$>$}
\def\<{$<$}

\def\ltsima{$\; \buildrel < \over \sim \;$}
\def\simlt{\lower.5ex\hbox{\ltsima}}
\def\gtsima{$\; \buildrel > \over \sim \;$}
\def\simgt{\lower.5ex\hbox{\gtsima}}

\begin{document}

   \thesaurus{           % A&A Section 08: Stars
              (08.05.1;  % Stars: early-type
               09.08.1)  % Interstellar medium: HII regions
             }
   \title{Time-dependent accretion and ejection 
implied by pre-stellar density profiles}

   \author{R. Henriksen$^{1,2}$, 
P.~Andr\'e$^{1}$\thanks{email: andre@sapvxg.saclay.cea.fr}, 
S.~Bontemps$^{1,3}$ }

   \offprints{P. Andr\'e}

   \institute{$^1$CEA, DSM, DAPNIA, Service d'Astrophysique, 
C.E. Saclay, F-91191 Gif-sur-Yvette Cedex, France \\
$^2$Department of Physics, Queen's University at Kingston, Ontario, Canada\\
$^3$Stockholm Observatory, S-133 36 Saltsj\"obaden, Sweden}

   \date{Received 5 August 1996 / Accepted 2 January 1997}

   \maketitle
\markboth {Henriksen et al.: Time-dependent accretion and ejection}{ }

   \begin{abstract} A recent homogeneous study of outflow activity 
in low-mass embedded young stellar objects
(YSOs) (Bontemps et al. 1996) suggests that mass 
ejection {\it and} mass accretion both decline significantly with time 
during protostellar evolution. 
In the present paper, we propose that 
this rapid decay of accretion/ejection activity 
is a direct result of the non-singular density 
profiles characterizing pre-collapse clouds. 
Submillimeter dust continuum mapping indicates 
that the radial profiles  of pre-stellar cores 
flatten out near their centers, being much flatter than 
$\rho(r) \propto r^{-2}$ at radii
less than a few thousand AU (Ward-Thompson et al. 1994). In some cases, 
sharp edges are observed at a finite core radius. 
Here we show, through Lagrangian analytical calculations,  
that the supersonic gravitational collapse of pre-stellar cloud cores 
with such centrally peaked, but flattened density profiles leads to 
a transitory phase of energetic accretion immediately following the 
formation of the central hydrostatic protostar. 
%hydrostatic stellar nucleus. 
Physically, the collapse occurs in various stages. 
%various collapse stages can be distinguished. 
The first stage corresponds to the nearly isothermal, dynamical collapse of 
the pre-stellar flat inner region, which ends with the formation of 
%collapses first nearly homologously to 
a finite-mass stellar nucleus. This phase is essentially non-existent 
in the `standard' singular model developed by Shu and co-workers.
%found by Shu (1977). 
In a second stage, the remaining cloud 
core material accretes supersonically onto a non-zero point mass. 
Because of the significant infall velocity field achieved 
during the first collapse stage, the accretion rate is initially higher 
than in the Shu model. This enhanced accretion 
persists as long as the gravitational pull of the initial point mass 
%, which does not exist in the Shu solution, 
remains significant. 
The accretion rate then quickly converges towards the characteristic value 
$\sim a^3/G$ (where $a$ is the sound speed), which is also the constant rate 
found by Shu (1977). 
If the model pre-stellar core has a finite outer boundary, there is  
a terminal decline of the accretion rate at late times due to the finite 
reservoir of mass.\\
We suggest that the initial epoch of vigorous accretion predicted by our 
non-singular model coincides with Class~$0$ protostars, 
which would explain their unusually powerful jets compared to the 
more evolved Class~I YSOs. 
We use a simple two-component power-law model to fit
the diagrams of outflow power versus envelope mass observed by 
Bontemps et al. (1996), and suggest that Taurus and $\rho$ Ophiuchi YSOs 
follow different accretion histories because of differing initial conditions. 
While the isolated Class~I sources of Taurus are relatively well explained by 
the standard Shu model, most of the Class~I objects of the $\rho$ Oph 
cluster may be effectively in their terminal accretion phase.

\end{abstract}
  

\section{Introduction}

Despite recent observational and theoretical progress, 
the initial conditions of star formation and the first phases of 
protostellar collapse remain poorly known. 

%According to current theoretical ideas, 
It is reasonably well established that low-mass stars form from 
%the inside-out collapse of dense 
the collapse of centrally-condensed cloud cores initially supported against 
gravity by a combination of thermal, magnetic, and turbulent pressures 
(e.g. Shu et al. 1987, 1993 for reviews). 
However, the critical conditions beyond which a 
cloud core becomes unstable and starts to collapse are uncertain and still 
a matter of debate (e.g. Shu 1977, Mouschovias 1991, Boss 1995, 
Whitworth et al. 1996). 
In particular, they
depend on yet unmeasured factors such as the strengths
of the static and fluctuating components of the magnetic field.\\
%However, 
Once fast cloud collapse sets in, 
the main {\it theoretical} features of the dynamical 
evolution that follows 
%the onset of fast cloud collapse 
have been known since the pioneering work
of Larson (1969). During a probably brief first phase, 
the released gravitational energy is freely radiated away and the cloud 
stays isothermal. This initial collapse phase,  
%proceeds in a very non-homologous fashion and 
which tends to produce a strong central concentration of matter,  
%with an $\rho \propto r^{-2}$ density gradient, 
ends with the formation of an opaque, hydrostatic stellar object 
(cf. Larson 1969; Boss \& Yorke 1995).
This time is often denoted $t = 0$ and 
referred to as (stellar) 
`core formation' in the literature (e.g. Hunter 1977). 
When the stellar core has fully formed,
one enters the {\it accretion phase}
during which the central protostar builds up its 
mass ($M_{\star}$) from a surrounding infalling envelope (of mass
$M_{env}$) while progressively warming up. 
The infalling gas is arrested and thermalized in an
accretion shock at the surface of the stellar core, generating an
infall luminosity L$_{inf} \approx GM_\star(t)\dot{M}_{acc}/R_\star$. 
In the `standard' theory of Shu et al. (1993) which uses 
singular isothermal spheres as initial conditions,
the accretion rate $\dot{M}_{acc}$ is constant and equal to 
$a_{eff}^3/G$, where $a_{eff}$ is the effective sound speed.
%Although idealized, the 
The singular isothermal sphere, which has 
$\rho \propto (a_{eff}^2/G)\, r^{-2}$, is 
a physically meaningful starting point for the 
`self-initiated' collapse of an isolated 
cloud core because it represents 
the (unstable) limit of infinite central concentration 
in equilibrium models for self-gravitating isothermal spheres (e.g. Shu 1977 -- 
hereafter Shu77).
Furthermore, Lizano \& Shu (1989) have shown that magnetically-supported 
cloud cores undergoing ambipolar diffusion evolve naturally toward a singular 
configuration reminiscent of a singular isothermal sphere (see also 
Ciolek \& Mouschovias 1994). However, it is likely 
that actual cloud cores become unstable and start to collapse before 
reaching the asymptotic singular state, 
especially when they are perturbed by an external 
agent such as a shock wave (e.g. Boss 1995, Whitworth et al. 1996).  
In this case, the initial conditions will be effectively non-singular,
%With non-singular, 
and $\dot{M}_{acc}$ is 
%predicted 
expected to be time-dependent (e.g. Zinnecker \& Tscharnuter 1984, 
Henriksen 1994, Foster \& Chevalier 1993, McLaughlin \& Pudritz 1997). 
It is this possibility that we explore further and compare with 
relevant observations in the present paper.\\ 
An unsettled related issue concerns the manner in which 
the central hydrostatic stellar core develops.  
While in some models  
core formation occurs in a dynamical,  
{\it supersonic} fashion (e.g. Larson 1969, Foster \& Chevalier 1993, 
present paper), 
this formation  is achieved  
by slow, {\it subsonic} evolution of the
gas in the scenario advocated by Shu and co-workers 
(see also Ciolek \& Mouschovias 1994).
An important observational consequence is that 
while the dynamical models predict the
existence of `isothermal protostars' 
(in the sense of the initial collapse phase outlined above), 
these {\it do not exist} in the standard Shu theory. 
As we will see, in practice 
both situations probably occur in nature.

Observationally, one distinguishes various empirical stages in the evolution 
of young stellar objects (YSOs) 
from cloud core to (low-mass) main sequence star (e.g. Lada 1987, 
Andr\'e 1994). The youngest observed YSOs 
are the Class~0 sources 
identified by Andr\'e, Ward-Thompson, \& Barsony (1993 -- hereafter AWB93),
which are characterized by very strong emission in the submillimeter 
continuum, virtually no emission below $\lambda \sim 10\ \mu$m, and 
powerful jet-like outflows. 
Their very high ratio of submillimeter to bolometric luminosity suggests 
they have $M_{env} >> M_{\star}$. Thus, 
Class~0 YSOs are excellent candidates for being very young protostars 
(estimated age $\sim 10^4$~yr) in which the hydrostatic core has
formed but not yet accreted the bulk of its final mass (AWB93). 
The next, still deeply embedded, YSO stage corresponds to the Class~I 
sources of Lada (1987), which are detected in the near-infrared 
($\lambda \sim 2\ \mu$m) and
have only moderate submillimeter continuum emission (Andr\'e \& Montmerle 
1994; hereafter AM94). They are interpreted as
more evolved protostars (typical age $\sim$ 10$^5$~yr) 
surrounded by both a disk and a residual circumstellar envelope 
of substellar mass ($\sim$ 0.1-0.3~$\sm$ at most in $\rho$ Ophiuchi; cf. AM94). 
Finally, the most evolved (Class~II and Class~III) YSO stages correspond to pre-main 
sequence stars (e.g., T~Tauri stars) 
%(``Classical'' and ``Weak'' T~Tauri stars, respectively) 
surrounded by 
a circumstellar disk (optically thick and optically thin at 
$\lambda \la 10\ \mu$m, 
respectively), but lacking a dense circumstellar envelope.\\ 
Many examples of pre-collapse, pre-stellar cloud cores 
are known. In particular, Myers and co-workers have 
%found and 
studied a large number of ammonia dense cores without $IRAS$ sources 
(e.g. Benson \& Myers 1989) which are traditionally associated with
(future) sites of {\it isolated} 
low-mass star formation (see Myers 1994 for a recent 
review). These starless dense cores, which are gravitationally bound 
and close to virial equilibrium, are believed to be magnetically supported and
to progressively evolve towards higher degrees of central concentration
through ambipolar diffusion (e.g. Mouschovias 1991).\\ 
More compact starless condensations 
%close to gravitational virial equilibrium
have been identified in regions of {\it multiple} 
star formation such as the 
$\rho$ Ophiuchi main cloud  
(e.g. Loren, Wootten, \& Wilking 1990, Mezger et al. 1992b, AWB93).  
In these regions, the weakness of the static magnetic field 
(Troland et al. 1996) and the complexity of its geometry 
%the magnetic field geometry 
%inferred from optical and near-infrared polarimetry proves to be complex
(Goodman \& Heiles 1994 and references therein)  
suggest that gravitational forces overpower magnetic ones and that an
external trigger rather than ambipolar diffusion 
is responsible for cloud fragmentation and core formation
%there 
%(e.g. Vrba 1977, Loren \& Wootten 1986).\\ 
(e.g. Loren \& Wootten 1986).

In the present paper, we bring together two independent sets of observational 
results recently obtained on the density structure of pre-stellar dense cores
(Sect. ~2.1) and on the evolution of protostellar outflows (Sect. ~2.2), and 
interpret them by means of a simple analytical theory (Sect. ~3.3)
which sheds light on the accretion histories found in 
%on some aspects of 
the numerical work of Foster \& Chevalier (1993 -- hereafter FC93).
%FC93.  
We compare our theoretical predictions with observations in Sect.~4 and 
we conclude in Sect. ~5.

\section{Relevant observations}

\subsection{Constraints on initial conditions: 
flattened pre-stellar density profiles}

Recent submillimetre dust continuum mapping
shows that the radial density profiles of pre-stellar cores are relatively 
steep towards their edges (i.e., sometimes steeper 
than $\rho(r) \propto r^{-2}$) but  
{\it flatten out} near their centres, becoming less steep than $\rho(r) \propto 
r^{-2}$ (Ward-Thompson et al. 1994 -- hereafter WSHA;  Andr\'e, Ward-Thompson, 
\& Motte 1996 -- hereafter AWM96; Ward-Thompson et al. 1997).
%The L1689B core in the $\rho$~Ophiuchi complex is the 
%best studied case so far. 
A representative, well-documented example 
of an isolated pre-stellar core is provided by  
L1689B, which is located in the $\rho$~Ophiuchi complex 
but outside the main cloud.
In this case, the radial density profile  
approaches $\rho(r) \propto r^{-2}$ between $\sim 4000$~AU and
$\sim 15000$~AU, and is as flat as $\rho(r) \propto r^{-0.4}$ 
or $\rho(r) \propto r^{-1.2}$ (depending on the deprojection hypothesis)
at radii less than $\sim 4000$~AU (AWM96). 
The mass and density of the relatively flat central region are estimated to be 
$\sim 0.33\ M_\odot$ and $\sim 2 \times 10^5$~cm$^{-3}$, respectively.\\ 
By contrast, {\it protostellar} envelopes are always found to be strongly centrally
peaked and do {\it not} exhibit the inner flattening seen in {\it pre-stellar} cores. 
In particular, isolated protostellar envelopes 
%in regions of isolated star-formation such as Taurus
have estimated radial density profiles which range
from $\rho(r) \propto r^{-1.5}$ to $\rho(r) \propto r^{-2}$ 
over more than 0.1~pc in radius (e.g., Ladd et al. 1991; Motte et al. 1996).
This difference of structure between the pre-collapse stage 
and the protostellar stage is illustrated in Fig.~1 which
compares the radial intensity profiles of the pre-stellar core L1689B 
and the candidate Class~0 object L1527.\\
We also note  that recent large-scale, high-angular resolution imaging
of the $\rho$ Ophiuchi cloud with the mid-infrared camera ISOCAM on board the 
ISO satellite (Abergel et al. 1996) 
suggests dense cores in clusters are often characterized by 
very sharp edges (i.e., steeper than $\rho \propto r^{-3}$ or 
$\rho \propto r^{-4}$), possibly produced by external pressure. (The 
$\rho$~Oph dense cores are seen as deep absorption structures 
by ISOCAM.)
\begin{figure}
\centerline{\hbox{\psfig{file=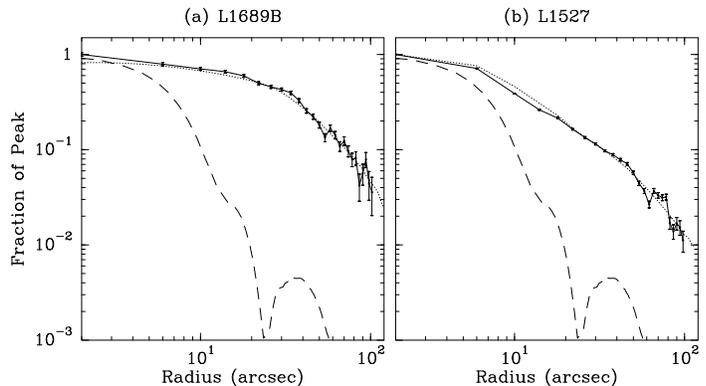,height= 6.5 cm,angle=270}}} 
\caption { Azimuthally averaged flux density profiles of the pre-stellar core
L1689B (a) and of the candidate Class~0 protostar L1527 (b) 
(adapted from AWM96 and Motte et al. 1996, respectively). 
Model profiles that fit the data are shown as dotted lines.
The L1689B model is isothermal and has 
$\rho(r) \propto r^{-0.4}$ for $r < 4000$~AU (i.e., $\theta < 25''$)
and $\rho(r) \propto r^{-2}$ for $r \geq 4000$~AU; the L1527 model 
has $T(r) \propto r^{-0.4}$ and $\rho(r) \propto r^{-1.5}$. These model profiles
result from a full simulation of the continuum dual-beam mapping technique
(see Motte et al. 1996), including a convolution with the beam of the telescope
which is shown as dashed lines. Note how flat the L1689B profile is compared 
with the L1527 profile for $\theta < 25''$. 
The apparent flattening of the L1527 profile 
for $\theta \simlt 5''$ is entirely attributable to the finite resolution of 
the telescope.}
\end {figure}

These results on the structure of dense cores 
set constraints on the initial conditions for gravitational collapse. 
The starless cores detected in the submillimeter continuum 
are thought to be descendants of cores without detectable 
submillimeter emission (which have lower densities) and
precursors of cores with embedded low-mass protostars. 
Based on the numbers of sources observed in the three groups of cores, 
the lifetime of starless cores with submillimeter emission was  
estimated to be $\sim 10^6$~yr by WSHA. So far  
a dozen pre-stellar cores have been studied in detail in the submillimeter, and 
an inner flattening of the radial density profile has been found for
{\it all} of them. Therefore, the 
%(presently poor) 
statistics indicate that  
the transition from flat to steep inner density profile occurs 
on a timescale shorter than $\sim 10^5$~yr, 
consistent with a phase of 
fast, possibly supersonic collapse (Ward-Thompson et al. 1997).\\ 
Ambipolar diffusion models of core formation do predict that
the time spent from the critical to the singular state is 
relatively short (typically $\simlt 10^6$~yr in the models of Mouschovias and
co-workers). 
Thus, in itself, the fact that most starless cores are observed in the 
more long-lived state when the central density profile is fairly flat does
not contradict these models. However, close comparison with the parameter 
study presented by Basu \& Mouschovias (1995) shows that the 
above timescale estimates agree roughly with the 
magnetically-supported core
models {\it only if} the initial cloud is highly subcritical 
(Ward-Thompson et al. 1997), which implies
rather high values of the (static) core magnetic field 
($\sim 80\ \mu$G in the case of L1689B -- see AWM96). 
Upper limits to the field in cloud cores obtained through Zeeman observations 
are typically lower than that 
(e.g. Crutcher et al. 1996, Troland et al. 1996),
%for the case of $\rho$~Oph), 
which suggests the observed 
dense cores cannot all be supported by a static magnetic field 
and cannot all have  
formed by ambipolar diffusion alone. 
The same conclusion is independently supported
by a statistical analysis of the observed core shapes which indicates that 
the overall morphology is prolate for most cores (Myers et al. 1991;  
Ryden 1996). This tends to go against ambipolar diffusion models which  
produce oblate cores, flattened in the direction 
of the mean magnetic field.\\ 
Therefore, although a more complete study of the structure of starless cores
would be desirable, 
%would be needed to draw definitive conclusions,  
we take the present observational evidence to suggest that, 
in some cases at least, 
%dense cores do not form by quasi-static ambipolar diffusion and 
the initial conditions for rapid protostellar collapse are characterized by a 
centrally flattened density profile and differ significantly from a 
singular isothermal sphere. 

\subsection{Decline of outflow power with time}

Class~0 protostars tend to drive highly collimated or 
``jet--like'' CO molecular outflows
(see review by Bachiller 1996). 
The mechanical luminosities of these outflows  
%approach $\sim 50$~\% of 
are often of the same order as 
the bolometric luminosities of the central sources (e.g., AWB93).
In contrast, 
while there is growing evidence that 
some outflow activity exists throughout the embedded phase,
%(e.g., Terebey et al. 1989; Parker et al. 1991), 
the CO outflows from Class~I sources tend to be poorly collimated  
and much less powerful than those from Class~0 sources. 

%
% fig2_acc.greg
%
\begin{figure}
\centerline{\hbox{\psfig{file=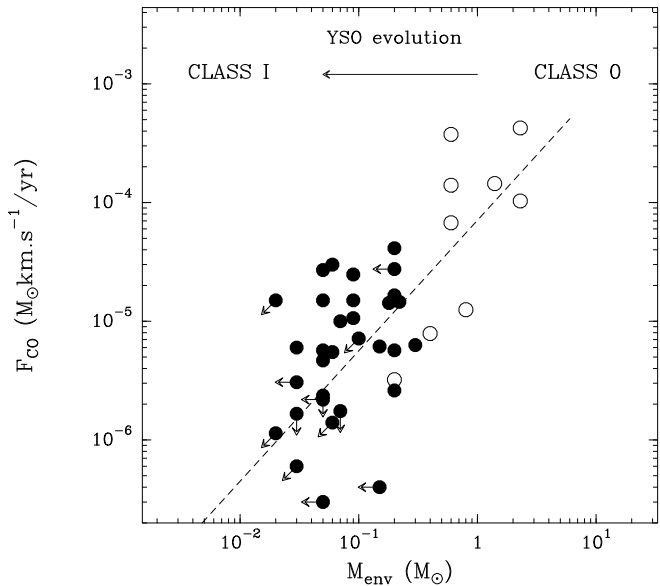,height=8cm,angle=270}}}
\vspace{8cm}
\caption{Outflow momentum flux $F_{\rm CO}$ %$\fco$  
versus circumstellar envelope mass $\protect\menv$ for a  
sample of nearby Class~I (filled circles) and Class~0 
(open circles) YSOs (Bontemps et al. 1996). 
(The arrows on the data points indicate upper limits.)
The `best fit' $\protect\fco$--$\protect\menv $ correlation  
is plotted as a dashed line. This diagram shows a clear decline of 
outflow power from Class~0s to Class~Is.}
\end {figure}

In an effort to quantify this evolution of molecular outflows during the 
protostellar phase, Bontemps et al. (1996 --  BATC) have recently obtained and
analyzed a homogeneous set of CO(2--1) data around a large sample of 
low-luminosity ($L_{bol} < 50\ L_\odot $), nearby ($d < 450$~pc) 
embedded YSOs, including 36 Class~I sources and 9 Class~0 sources.
The results show that 
essentially {\it all} embedded YSOs have some degree of outflow
activity, suggesting the outflow phase and the infall/accretion phase  
coincide. This is consistent with the idea that accretion 
cannot proceed without ejection and that outflows are directly powered by 
accretion (e.g., Ferreira \& Pelletier 1995).\\ 
In addition, Class~0 objects are found to lie an order of magnitude 
above the well-known 
correlation between outflow momentum flux and bolometric luminosity
holding for Class~I sources. This confirms that 
Class~0 objects differ qualitatively from
Class~I sources, independently of inclination effects.   
On the other hand, as can be seen in Fig.~2, outflow momentum flux is roughly 
proportional to circumstellar envelope mass in the {\it entire} 
BATC sample (i.e., including both Class~I and Class~0 sources). 
%As illustrated in Figure~2,
Bontemps et al. argue that  
this new correlation is independent of the   
$\fco$--$\lbol$ correlation 
and most likely results from a progressive  
decrease of outflow power with time during the accretion phase. 

Since many theoretical models of bipolar outflows 
(e.g. Shu et al. 1994, Ferreira \& Pelletier 1995, Fiege \& Henriksen 1996a,b) 
predict a direct proportionality between accretion and ejection,  
 BATC further suggest that the observed decline
in outflow power 
reflects a corresponding decrease in the mass accretion/infall rate:  
In this view, $\macc$ would decline from $\sim 10^{-5}\,\myr$ 
for the youngest Class~0 protostars to $\sim 10^{-7}\,\myr$ for 
the most evolved Class~I sources and most active T Tauri stars.
 
\section{Relevant theory} 

We propose that the decrease of accretion rate seen by 
BATC during the protostellar phase 
(see Sect.~2.2 above) is a direct consequence of 
the non-singular initial conditions indicated by the 
pre-stellar core observations summarized in Sect.~2.1. 
Our fundamental thesis in this article is that  supersonic core formation 
and a subsequent core accretion flow may be adequately 
described as a zero pressure 
flow that evolves from a more general density profile than the $r^{-2}$ profile 
at a pre-core formation epoch ($t<0$).  
We use analytical methods to elucidate 
the physical origin of a sharp 
decrease of $\macc$ during the early accretion phase  
(i.e., just after the completion of stellar core formation). 
We show that 
the key to such a decrease is the existence of a transition 
between a flat inner region  
and an outer inverse square `envelope' in 
the initial density profile: the inner region collapses first to form a
hydrostatic stellar core, and the remaining cloud material then 
accretes supersonically onto a non-zero point mass.

\subsection{Overview of known spherical collapse solutions}

Here, we assemble some relevant non-linear studies 
of isothermal protostellar collapse in spherical geometry. 
The isothermality seems justified physically at early (e.g., pre-stellar) stages
(note this does {\it not} imply the dynamical importance of pressure), 
but the assumption of spherical symmetry implies the neglect of rotation, 
magnetic fields and dynamical turbulence.  

Mathematically, the 
%remarkable 
paper by Whitworth \&  Summers (1985 -- WS85) concludes 
%This paper gives the logical `denouement' to 
the series of original studies on isothermal,  
self-similar accretion flows
by Larson (1969), Penston (1969), Shu77 and Hunter (1977).
However physically the debate continues, based essentially on the 
failure of either 
the Larson-Penston solution or the Shu solution to describe 
the accretion rate as found 
in numerical simulations (e.g. Hunter 1977, FC93) 
after the core-formation epoch. 
%from the core-formation epoch to late times. 
The Shu solution describes the asymptotic behaviour satisfactorily, 
while the Larson-Penston solution seems more relevant near core formation 
although it is not perfectly so. 

The major contribution of WS85 
was to show that a parametrically stable infinity of `acceptable' 
%sound wave dominated 
similarity solutions\footnote{ Similarity solutions are functions 
defined by 
ordinary differential equations which nevertheless represent 
solutions 
to the full partial differential system of the physical problem. 
They also have the
advantage of removing initial and boundary conditions to infinity.
The physical justification for their use is the empirical fact 
that they arise as `intermediate asymptotes'  
in various regimes of more complicated flows 
(Barenblatt \& Zel`dovich 1982).}
exist rather than merely the previously known Larson-Penston (L/P) 
%homologous 
infall and the Shu (S) implosion behind an outward going expansion wave. 
`Acceptable' in this context implies that the solutions have reasonable 
initial conditions inside an inward going compression wave in general 
(that may 
be initiated by an external disturbance) and 
describe supersonic infall after the passage of this wave. True free-fall 
accretion is established only after this wave has reached the centre of the system 
at $t=0$. It develops behind the rebounding expansion wave.
%outward propagating rebonding wave.
 
WS85 classified all possible isothermal 
similarity solutions in a bounded two-dimensional space. 
Briefly, this continuum of solutions spans a wide range from 
`gravity dominated' 
solutions (corresponding to the `band~0/band~1' of WS85) to `sound-wave'  
dominated solutions (corresponding to high `band' numbers in WS85).
The sound-wave (or pressure) dominated solutions are those usually studied, 
which depend on the classical similarity variable $x = r/at$ (e.g., Shu77). 
The S solution is itself a limiting sound-wave dominated case 
(`band $\infty$' in WS85) 
that develops from an initial high density singular isothermal 
equilibrium. 
%(having a density profile $\propto r^{-2}$ everywhere).

One might think that the gravity dominated solutions should be determined 
in general by the 
form of the density or 
mass profile at some epoch 
in the region of supersonic flow, rather than by the isothermality as in 
the above family. We propose  
%the existence of such solutions signals
a more general type of similarity (see 
Appendix A and below) that arises from a given density profile in a region 
of supersonic flow. 
This idea  was also hinted at in Hunter (1977) 
in the course of his explanation of the absence of shocks 
in the supersonic region prior to the arrival of the compression 
wave at $r = 0$. Hunter observes that gravity provides a non-propagating local 
acceleration by `action at a distance'. Formally, the flow ceases 
to be hyperbolic in its causal structure, becoming elliptic instead.

Such a dynamical change in self-similar symmetry was considered by 
Henriksen (1994 -- hereafter H94) 
to argue, 
on the basis of pressure-free, analytical calculations that 
non-constant accretion rates might be expected.   
We shall exploit this idea in the present paper after an 
assessment of the pressure dominated self-similar solutions.   
 
It is difficult to isolate precisely the epoch at which the change in self-similarity 
type occurs. We know from the simulations by Hunter (1977) 
and recently by FC93
that classical self-similarity arises in the course of the collapse. 
%This behaviour 
The flow  
is generally found to agree best with the L/P type solution, 
but the agreement is not 
perfect, since the accretion rate found by these numerical studies is  
{\it not} the constant value to be expected 
if the self-similar solutions 
persist at the supersonic core formation epoch ($t=0$). 
In fact the accretion rate passes through a peak and then settles only 
gradually (if the cloud is sufficiently large) to the Shu 
value (see Fig.~3 of FC93).
The nature of 
this transitory peak was left somewhat unclear by FC93 due to their problems 
of numerical resolution at $t=0$, but we believe its nature to be vital to 
the understanding of the outflow observations summarized in Sect.~2.2. 

\subsection{Justification of our pressure-free analytical approach}

\subsubsection{Neglect of magnetic field and rotation}

We neglect both magnetic field and rotation in our discussion. 
Mouschovias and co-workers 
(Mouschovias 1995 and references therein) have demonstrated 
that a magnetically-supported, subcritical cloud evolves naturally to a 
centrally peaked, flattened density profile during a quasi-static phase of 
ambipolar diffusion. This mechanism can successfully account for the 
formation and structure of some isolated pre-stellar cores. 
%(see Sect.~2.1 above).
%This is not in conflict with the discussion above and below. 
The quasi-static phase ends with the formation of a magnetically 
supercritical core that collapses in a dynamical but 
magnetically-controlled fashion, while the bulk 
of the cloud remains magnetically supported. 
In Mouschovias' models, the `dynamical'  
contraction of the supercritical core is almost always subsonic and 
evolves only gradually towards free fall, which is not attained up to 
densities
as large as $\sim 10^9$--$10^{10}$~cm$^{-3} $
(see Basu \& Mouschovias 1995 and Fig.~2d of Ciolek \& Mouschovias 1994). 
This corresponds to a rather gentle process of 
hydrostatic stellar core formation taking significantly longer than 
a free-fall time, 
i.e., typically on the order of $10^6$~yr in published models 
from the time the parent pre-stellar core becomes
magnetically supercritical at a central density $\sim 10^5$~cm$^{-3} $.\\ 
However, as pointed out in Sect.~1 and Sect.~2.1, some cloud cores are 
apparently 
not magnetically supported and appear to evolve faster than predicted by 
ambipolar diffusion models. 
%In fact we note that, even if clouds are initially magnetically supported, 
Furthermore, even in the case of initial magnetic support, 
superalfv\'enic implosion of cloud fragments is possible, 
for instance induced by a shock (e.g. Mouschovias 1989,  
Tomisaka, Ikeuchi, \& Nakamura 1988, Tomisaka 1996). 
This may well be a dominant process in regions 
of multiple star formation.  
Thus, we believe the collapse can become vigorous and 
supersonic (superalfv\'enic as the case may be) relatively early on, at least 
in a small central mass (see the next section).  
This is the 
hypothesis we will make in the remainder of the paper. 

Rotation and magnetic fields are thought to be essential to the 
launching of the bipolar outflows that are invariably 
found to be associated with 
protostars (e.g. K\"onigl \& Ruden 1993, Henriksen \& Valls-Gabaud 1994).  
This argues in favour of their importance, at least when the outer parts 
of the initial cloud core fall in.
%(see also Galli \& Shu 1993). 
It is likely that they are also 
important to the formation and evolution of the flattened inner parts.
%core material.

Thus we cannot expect to include 
the whole picture self-consistently in our subsequent arguments which ignore
magnetic fields.
However if gravity is the main 
player in the core after a `point of no return' is passed, then our present approach remains crudely justified, if incomplete.

\subsubsection{Lagrangian treatment of Isothermal Self-Similarity}

As discussed by Henriksen (1989 -- H89), spherically symmetric, isothermal, 
self-similar collapse can be very conveniently treated when one uses a Lagrangian formalism. 
This is recalled briefly in Appendix~A for the simplest cases. 
Here, we merely want to show that such a formalism
provides continuity with the supersonic treatment to be found in Sect.~3.3. 
The advantage of the Lagrangian formulation is that self-similar solutions are
fully described by the generalized `Friedmann, Lema\^\i tre, Robertson-Walker' 
(FLRW) differential equation (Eq.~A2 in the limit of Eqs.~A4 and A5):
 
$$\left({dS\over d\xi}\right)^2-{1\over S} + \alpha_s^2H=k, \eqno(1)$$
which expresses the conservation of energy.
In this equation, $\xi\equiv \sqrt{2GM(r)/r^3}\,(t-  t_o)$ is 
the self-similar variable, where $M(r)$ is the mass distribution 
(i.e., the mass contained within the sphere of radius $r$ as a function of 
$r$) at some reference epoch $t = t_o$ (the reason for this time 
offset will become apparent later).
The function $S(\xi)\equiv R(t,r)/r$ gives the normalized 
%comoving (or Lagrangian) coordinate 
position at time $t$ of the shell that was initially 
($t =  t_o$) at $r$. 
Finally, $\alpha_s^2\equiv a_s^2 r/GM(r)$ where $a_s$ is the 
isothermal sound speed, 
$H=\ln{\rho/\rho_o}$ where $\rho_o(r)$ is 
the density profile at $t=  t_o$, and 
$k \equiv E(r)r/GM(r) = \pm 1$ or 0 is 
the normalized total energy ($E(r)$ being the specific total energy 
of the shell
initially at $r$). 
The solutions describing a collapse from rest have $k=-1$, and those with  
zero total energy have $k=0$ (we omit other cases for simplicity). 

The present formulation is unique in that it allows 
for a transition to a purely self-gravitating flow 
wherein the pressure is negligible 
(formally $\alpha_s\rightarrow 0$) {\it before} the ingoing compression wave 
establishes the $r^{-2}$ `attractor' at $t=0$ 
[Blottiau, Bouquet, \& Chi\`eze (1988) were the first 
to use this terminology]. In order that $\alpha_s^2\rightarrow 0$ 
at $t<0$ in an isothermal flow, it is clear that $M(r)/r$ must tend to 
a large value as $r\rightarrow r_N$, where $r_N$ is the Lagrangian radius 
of this core or nuclear region. This can happen {\it if a sufficiently 
dense mass with a flattened density profile 
forms at the centre of the system at $t<0$}. 
This region should eventually 
collapse nearly homologously to form a central point mass
so long as the inflow is 
supersonic (or super fast-Alfv\'enic as the case may be) and pressure 
support is negligible.
 
Assuming that some fraction 
of the core does collapse in the free-fall time $t_{ff}(N)$, 
we begin our zero pressure calculations  
at the onset of this rapid collapse. We define this starting point 
as being $  t_o = - t_{ff}(N)$, for consistency with the standard 
pressure wave connected solutions. (With this definition, 
stellar core formation is complete at $t=0$, which also marks the beginning of 
the accretion phase.)  
When the parameter $\alpha_s^2\rightarrow 0$,
one has made the transition of course from an hyperbolic system to an elliptic 
one. In other words, there is a wave-mitigated causality when pressure is 
important, and action at a distance (in Newtonian gravity) 
in the zero pressure limit (see also Hunter 1977). 
This is discussed briefly in Appendix A.   

\subsubsection{The case for starting from strictly flat inner density profiles} 

The density profiles observed in pre-stellar clouds (Sect.~2.1) 
are not as centrally flat as are some of the self-similar solutions of WS85, 
nor indeed of the magneto-static cores of Basu \& Mouschovias (1995). 
This is not very surprising since 
a flat profile indicates a relatively low pressure gradient so that the 
hydrostatic equilibrium required for long lived clouds is not possible. In 
this state the cloud is in fact approaching the super-critical state.  
The real question then is whether the collapse proceeds in general by some 
part of the material passing 
through a phase of nearly zero 
%central density or 
pressure gradient and collapsing homologously, 
as the band~0 sound wave dominated self-similar 
solutions suggest (see Figs. 5, 6, 7 of WS85). Interestingly enough, 
the numerical calculations by FC93 that begin from non self-similar 
conditions show a velocity distribution in the critical Bonnor-Ebert sphere 
at early times that would tend to {\it flatten} the density profile 
(see e.g. the 
discussion in the second paragraph of their Sect.~2.3) 
while mildly sub-critical 
cases show oscillations as are found in the \lq bound' (band 1 and higher) self-similar solutions. FC93  
interpret their maximum velocity peak in terms of a flat interior density 
distribution that passes abruptly to a $r^{-2}$ profile.  
In fact, it is easy to show by linearizing the equations  
that an initially flattened density profile will first tend to 
evolve towards a still more flattened configuration on a free-fall timescale 
set by the cloud central density. 
Physically this is due once again to the velocity profile of the fastest 
growing linear mode that is  consistent with zero velocity at the 
cloud edge and at the cloud centre, since this is such that 
the outer layers overtake the inner ones. The linear calculation can 
only be suggestive, but in the generalized solutions of Blottiau et al. (1988) 
as well as in the band~0 solutions of WS85, flattened centres and shells 
(which would also collapse to give a point mass) appear. Moreover during 
the `quasi-static' evolution followed by Basu \& Mouschovias (1995), 
flattened supercritical cores appear. That a portion of this core collapses 
homologously is however less clear from their calculations (depending on the 
ion-neutral collision time). 

\subsection{Accretion phase history} 

For the reasons given in Sect.~3.2 above, 
because of the nature of the abundant non-linear 
self-similar solutions (WS85), and because of the suggestive numerical study of 
FC93, we now proceed to calculate an accretion history 
for a self-similar cloud based on the following 
hypothesis of pre-collapse evolution.\\ 
In agreement with the observations, we suppose that the 
pre-collapse cloud core evolves quasi-statically, and develops a dense central region 
with a nearly flat density profile, until an external disturbance initiates
the collapse.
%until it contains a central region with a nearly flat density profile. 
We further suppose that, after a transitional period 
(leading to, e.g., magnetic decoupling), 
the flat region is free to collapse suddenly, essentially homologously
due to the absence of a pressure gradient. 
It is possible that further flattening of the profile occurs during the collapse. 
In any case the beginning of 
this free collapse is taken once again to be at  
$t =  t_o \equiv -t_{ff}(N)$,
so that the completion of the core collapse is  
at $t=0$, where $t_{ff}(N)$ is
the free-fall time of a uniform sphere from rest, defined explicitly below.  
For all times $t > t_o$ we shall assume that the  
evolution is of the gravity-dominated self-similar form, 
given uniquely by the density profile at the $t = t_o$ epoch (see Appendix A). 
This is all that is required for the purposes of our calculation, 
but the flow has probably ceased to be pressure dominated even before  
$t= t_o$.\\ 
An important point to realize is that the accretion rate at $t=0$ 
will not yet be that appropriate to the exterior of the cloud, since this will be for some time dominated by the collapsed core mass rather than by the self-gravity appropriate to the outer density profile.
The subsequent transition to a constant asymptotic accretion rate happens as 
sufficient mass from the outer cloud is accreted to dominate the initial 
core mass.

\subsubsection{\it General Formulation}

The equations include Eq.~(1) above, plus those of 
Appendix A and of H89, H94 but with 
$\alpha_s^2=0$.  
We carry out the calculations for a family of 
initial density profiles which are all strictly flat for 
$r \le r_N$ (see Fig.~3 below). 
The density is measured in units of the density $\rho_N$ of 
the flattened nucleus of the pre-collapse cloud 
(i.e., $\tilde{\rho} \equiv \rho /\rho_N$), 
the mass is in units of the mass nucleus
$M_N$ (i.e., $\tilde{M} \equiv M/M_N$) 
and the radius is in units of $r_N$ (i.e., $\tilde{r} \equiv r/r_N$), 
where $r_N$ is the radius of the nuclear region just before free-fall
collapse. For convenience, 
time is measured from the beginning of the free collapse 
($t = t_o = -t_{ff}(N)$) 
in units of the nuclear region free-fall time 
$t_{ff}(N) = (3\pi/32G\rho_N)^{1/2} $, that is we introduce the dimensionless variable 
%$$\tau \equiv {2\over\pi}\sqrt{{8\pi G\rho_N\over 3}}\, (t- t_o).\eqno(2)$$ 
$$\tau \equiv {t- t_o \over t_{ff}(N)}.\eqno(2)$$
The stellar core is fully formed at $\tau=1$
%$\tau=0$ 
and consists of  
all shells with Lagrangian labels $\tilde{r} \le 1$. 
We take this epoch to correspond to $\xi =\pi/2$ since 
this is appropriate to the collapse of a uniform sphere from rest 
(e.g. Hunter 1962). In this description, $\tau=1$ also marks the 
beginning of the YSO accretion phase, while dimensionless times in the range 
$0 < \tau < 1$ correspond to `isothermal protostars' (i.e., collapsing 
cloud fragments with no central hydrostatic core -- see Sect.~1).\\
The introduction of the self-similar variable $\xi$ ($0 \leq \xi \leq \pi/2$) 
is partly justified by the fact that 
the free-fall time of the shell with Lagrangian label $\tilde{r}$ 
is related to the mass function $M(r)$ through

$$t_{ff}(\tilde{r})  = 
\sqrt{\tilde{r}^3 \over \tilde{M}(\tilde{r})} \, t_{ff}(N) 
= ({\pi/2 \over \xi }) \, (t-  t_o) . $$ 
%= ({\pi/2 \over \xi }) \, (t + t_{ff}(N)) . $$
%In addition, from the definition of $\xi$ ($\xi\equiv \sqrt{2GM(r)/r^3}\,t$), 
Thus, for $\tilde{r}\ge 1$ we have $\xi \leq \pi/2$ and 

$${\xi \over \pi/2} = 
\tau \, \sqrt{\tilde{M}(\tilde{r}) \over \tilde{r}^3}.\eqno(3)$$

(In the post nuclear formation epoch,
we have $\xi =\pi/2$ for all $\tilde{r}\le 1$ since these shells 
have already collapsed in the singular nucleus.) 

We shall be interested in those shells which are accreting near the center 
during the post nuclear formation phase and thus which are about to enter the singularity. 
As is well known in cosmology (FLRW models, see H89), for such 
shells a collapse in the absence of significant pressure gradients can be described 
by Eq. (1) with $k=0$ (and $\alpha_s^2=0$), 
for which the solution is 
$$S(\xi) =\left({3\over 2}({\pi\over 2}-\xi)\right)^{2/3},\eqno(4)$$
for $\xi\le \pi/2$. 

We note immediately that $\tau=1$ when $\xi=\pi/2$ for the outer shell of 
the nucleus ($\tilde{r} =1$), which is the instant when the nucleus has fully formed. 

To obtain the relationship between the real position of a spherical shell 
$R(\tau,r)$ and its Lagrangian or comoving label $r$ we must use $R = rS(\xi)$, 
which with Eq. (4) gives 
$$ {\xi\over\pi/2}= -{4\over 3\pi}\left({R\over r}\right)^{3/2}+1.\eqno(5)$$
We see that $R=0$ when $\xi=\pi/2$ as expected. 
Note however that nevertheless the shell label $\tilde{r}$ 
will be greater than or equal to $1$ in our calculations, since we shall be 
studying spheres whose initial radii are greater than or equal to $r_N$.

Our goal is to calculate the accretion rate history and  the 
mass remaining in the envelope surrounding the collapsed nucleus, 
since both of these 
quantities are susceptible to observation. The accretion rate $\macc$  
is given at each spherical radius $R$ and time $t$ by 
$\macc =4\pi R^2\, \rho \, \partial_t R$. 
If we use our various definitions and measure distance and density in units as above, 
and velocity in units of the nuclear free-fall velocity $v_N\equiv\sqrt{2GM_N/r_N}$ 
we obtain 
$$\dot{\tilde{M}}_{acc} \equiv 
{\macc\over 4\pi r_N^2\rho_N v_N}=\tilde{r}^2\tilde{\rho} S^2 \, {dS\over d\xi}\, \sqrt{\tilde{M}(\tilde{r})\over \tilde{r}}, \eqno(6)$$
where $S$ is given by Eq.~(4) and $\tilde{\rho}$ follows from 
Eq.~(A3) in the form 
$$\tilde{\rho} ={\tilde{\rho}_o(r)\over S^3(1+{d\ln S\over d\ln \xi}
{\partial\ln\xi\over\partial 
\ln r})}, $$
where $\rho_o$ is the density profile at 
%$t=0$.
$t= t_o$.

Since we are mainly interested in calculating the 
accretion rate near the centre of the post-collapse configuration where 
$R/r\rightarrow 0$, we may let $\xi $ tend towards 
$\xi=\pi/2$ in Eq.~(6). 

%We also note from the definition of $\xi$ that, 
The mass $M_\star$ of the central collapsed object accreting the 
shell labelled $\tilde{r}$ at the time $\tau$ is also easily obtained 
by setting $\xi = \pi/2$ in Eq.~(3) :
$$\tilde{M}_\star(\tau)={\tilde{r}^3(\tau)\over \tau^2}.\eqno(7)$$
It is perhaps worth repeating 
this equation in the form that it assumes with restored dimensions, namely
$$M_\star (t)= {\pi^2\over 8}~{r^3(t)\over G (t- t_o)^2}.\eqno(7')$$ 
 
%This expression also gives 
Conversely, the time at which the shell initially at 
$\tilde{r}$ accretes onto the central singularity is 
$$\tau = \sqrt{{\tilde{r}^3 \over \tilde{M}(\tilde{r})}}.\eqno(8)$$
As for the total mass of the accreting circumstellar envelope 
at time $\tau$, it is simply  
$\tilde{M}_{env}(\tau) \equiv \tilde{M}_{cloud}-\tilde{M}_\star(\tau)$. 

\subsubsection{\it Two-component Power-Law Model}

To obtain explicit expressions for $\macc$ and $M_{env}$, we must assume 
an explicit initial density profile. We use a multi-component 
profile of the form (in the units introduced in Sect.~3.3.1) 
$$\tilde{\rho_o}= \tilde{r}^{-2/D_1},~~1 \le \tilde{r} 
< \tilde{r}_b,\eqno(9)$$
$$\tilde{\rho_o}= \tilde{r_b}^{-2/D_1}\, 
({\tilde{r} \over \tilde{r_b}})^{-2/D_2}
,~~ \tilde{r} \ge \tilde{r_b},\eqno(9')$$

and $\tilde{\rho_o}=1$ for $\tilde{r} \le 1$, 
where $\tilde{r}_b$ is a characteristic boundary radius,  
$D_1$ is a positive number characterizing the power-law slope of the cloud 
`envelope', and $0 < D_2 < 2/3$ characterizes the radial profile of the 
`edge' of the model pre-stellar core 
%$\not= 2/3$ 
(see Fig.~3). 
This form gives a weak discontinuity in the density profile 
between the flat nucleus and the $r^{-2/D_1}$ envelope
(the density remains continuous but its derivative is discontinuous at
$r = r_N$). 
This discontinuity breaks the {\it density} self-similarity but not the dynamical self-similarity as discussed in H89.
In a pressure-free epoch 
there is no reason why such a discontinuity might not be 
preserved, if ever it were to be established by an inward propagating wave 
for example. Such weak discontinuities often arise 
during the adjustment of a gas to a change 
in the boundary conditions (e.g. Landau \& Lifshitz 1987) 
and this may well be the manner in which the collapse is initiated. 
In this subsection, we will keep $D_1$ and $D_2$ as free parameters in order
to obtain general formulae and model a wide range of accretion histories, but 
in practice we will use $D_1 = 1$ ($r^{-2}$ envelope) and $D_2 = 0.1$ when 
comparing with observations (Sect.~4). The rationale behind such a 
two-component, idealized density profile is that it approximates the 
observed structure of pre-stellar cores reasonably well (cf. WSHA, AWM96, 
Abergel et al. 1996, and
Sect. 2.1) and can account for the necessarily finite radius of influence 
of protostars forming in clusters (since $D_2 < 2/3$ 
the model pre-stellar core has a steep edge and a finite total mass, 
see below). 
%
%  rho_dick.greg
\begin{figure} 
\centerline{\hbox{\psfig{file=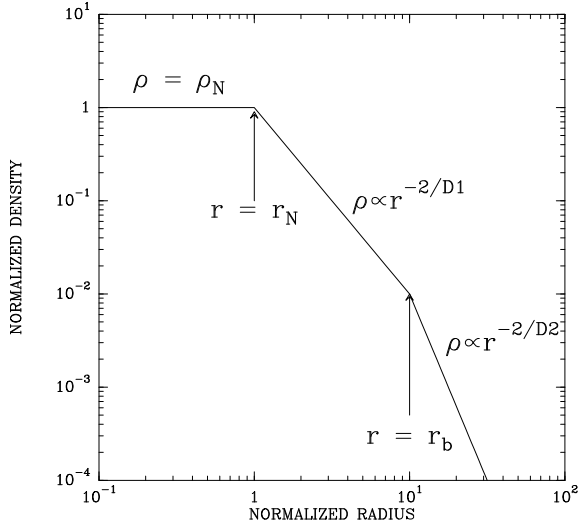,height= 8 cm,angle=270}}} 
%\vspace{6cm}
\caption {Density $vs.$ radius (in log--log format) for an idealized 
pre-stellar cloud with a flat nucleus surrounded by a power-law envelope 
and a steeper power-law outer boundary  
%Here, $D_1 = 1$ and $D_2 = 0.5$ were adopted
}
\end {figure}
With this density profile, the mass function $M(r)$ that appears in 
the self-similar variable $\xi$ (see Appendix A and H89) in the 
post `nuclear collapse' domain ($\xi>0$) is $\int_{r_N}^r\,\rho_o dr +M_N$, 
where the nuclear mass is $M_N = (4\pi/3)\rho_N r_N^3$. 
Thus, the mass distribution 
in units of $M_N$ takes the simple form 

$$\tilde{M}(\tilde{r})= (3D_1\tilde{r}^{(3-2/D_1)}-2)/(3D_1-2),
~~~~1 \le \tilde{r} < \tilde{r}_b, \eqno(10)$$ 
$$\tilde{M}(\tilde{r})= \tilde{M}_b + 
{3D_2\tilde{r}_b^{(3-2/D_1)} \over 3D_2-2}\, 
[({\tilde{r} \over \tilde{r}_b})^{(3-2/D_2)}-1]
,~~~~ \tilde{r} \ge \tilde{r_b},\eqno(10')$$

where $\tilde{M}_b = (3D_1\tilde{r}_b^{(3-2/D_1)}-2)/(3D_1-2)$ is the mass
enclosed within the boundary radius $\tilde{r}_b$.
Since $D_2 < 2/3$, 
the mass function converges at large $\tilde{r}$ towards
$\tilde{M}_{cloud} = \tilde{M}_b + [3D_2\tilde{r}_b^{(3-2/D_1)}/(2-3D_2)]$.
(Note that the case $D_2 = 2/3$, which we do not include explicitly here, 
is logarithmic.)
%when $D_2 < 2/3$ (which will be the case in practice).

In this case, 
the time function $\tau(\tilde{r})$ giving the time of arrival in the central
hydrostatic core of the shell 
initially at $\tilde{r}$ (see Eq.~8) can be easily inverted and 
has simple power-law asymptotic forms:
$\tilde{r}(\tau) \sim ({3D_1 \over 3D_1-2})^{D_1/2} \times \tau^{D_1}$
before accretion of the shell labeled $\tilde{r}_b$, and\\
$\tilde{r}(\tau) \sim \tilde{M}_{cloud}^{1/3} \times \tau^{2/3}$ 
at later times. 
Figure~4 displays this $\tilde{r}(\tau)$ relationship for three initial 
density profiles which all have a sharp outer boundary
($\tilde{r}_b = 10$ and $D_2 = 0.1$).
%
%  radtime_fig.greg
\begin{figure}
\centerline{\hbox{\psfig{file=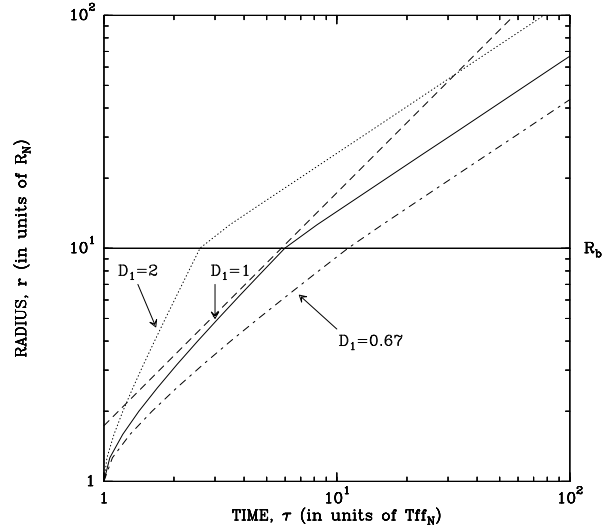,height= 8 cm,angle=270}}} 
%\vspace{8cm}
\caption {Radius of the shell accreting onto the central singularity as 
a function of time for three values of $D_1$ corresponding 
to $r^{-1}$ (dotted curve), 
$r^{-2}$ (solid curve), and $r^{-3}$ (dash-dotted curve)
envelopes respectively in the initial
density profile. A sharp outer edge ($\tilde{r}_b = 10$ and $D_2 = 0.1$) 
has been assumed in all three cases. The dashed line shows the linear 
relation $\tilde{r} = 3^{1/2} \tau$ expected for a singular $r^{-2}$ 
sphere. By definition, $\tau =1$ corresponds to the time at which the outer
shell of the flat nucleus reaches the central singularity}
\end {figure}
The special $D_1 = 1$ case ($r^{-2}$ `envelope', 
which we will use in practice) is particularly
instructive since it shows that a singular $r^{-2}$ pre-stellar core 
would collapse according to the {\it linear} relation 
$\tilde{r} = \sqrt{3}\, \tau$ or 
$r(t) = {2\sqrt{3}\over \pi}\, v_N \, (t- t_o)$ in restored 
dimensions. This makes possible a direct comparison 
with Shu77's expansion wave solution which predicts
$r(t) = R_{inf}/2  = (a_s/2)\, t$ (here $R_{inf}$ 
denotes the head of the expansion wave at time~$t$; note that 
$ t_o = 0$ in Shu's solution). 
In a sense, Fig.~4 illustrates that, {\it during the accretion phase}, 
the idealized solutions explored in this paper 
are characterized by an `inside-out' collapse behind an 
expansion wave reminiscent of the Shu solution. 
 
%
% macctime_fig.greg
\begin{figure}
\centerline{\hbox{\psfig{file=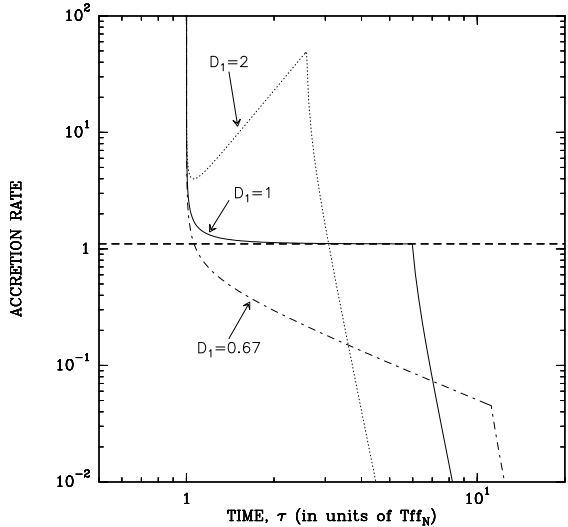,height= 8 cm,angle=270}}} 
%\vspace{8cm}
\caption {Accretion rate as a function of time 
for three values of $D_1$ corresponding 
to $r^{-1}$ (dotted curve), $r^{-2}$ (solid curve), 
and $r^{-3}$ (dash-dotted curve) envelopes respectively in the initial
density profile. A sharp outer edge ($\tilde{r}_b = 10$ and $D_2 = 0.1$) 
has been assumed in all three cases. The dashed horizontal line gives the asymptotic value of the accretion rate 
obtained in the case an $r^{-2}$ initial density profile }
\end {figure}
Substituting Eqs.~(10) and (10') into Eq~(6), we
obtain the desired accretion rate history at $\xi=\pi/2$ explicitly as
 
$$\dot{\tilde{M}}_{acc} =-({2\over 3\pi}){\tilde{r}^{(3-2/D_1)} \over \tau}\, {3D_1\tilde{r}^{(3-2/D_1)}-2 
\over \tilde{r}^{(3-2/D_1)}-1},\eqno(11)$$    

as long as the expansion wave is still within $\tilde{r}_b$. At later times, 
the accretion history may be obtained from the 
relation: 

$$\dot{\tilde{M}}_{acc} = -({4\over 3\pi}){\tilde{r}_b^{2/\bar{D}}
\tilde{r}^{(3-2/D_2)} \over \tau}\, {\tilde{M}(\tilde{r}) 
\over \tilde{M}(\tilde{r}) - 
\tilde{r}_b^{2/\bar{D}} \tilde{r}^{(3-2/D_2)} }, \eqno(11')$$
where $\bar{D}$ is defined by $1/\bar{D} \equiv (1/D_2) - (1/D_1)$.
Based on these formulae, Fig.~5 displays the 
predicted accretion rate as a function 
of time for the same three model pre-stellar profiles as in Fig.~4.
%
%   maccmenv_fig.greg
\begin{figure}
\centerline{\hbox{\psfig{file=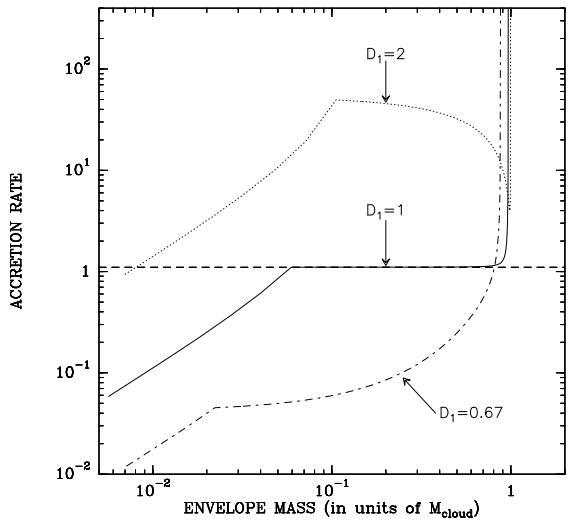,height= 8 cm,angle=270}}} 
%\vspace{8cm}
\caption {Absolute value of the accretion rate as a function of envelope mass  
for the same three initial density profiles as in Fig.~4 and Fig.~5. 
Here again, the dashed horizontal line gives the asymptotic value 
of the accretion rate 
predicted in the case an $r^{-2}$ initial density profile }
\end {figure}
The corresponding relationships between $\dot{\tilde{M}}_{acc}$ 
and $\tilde{M}_{env}$ are shown in Fig.~6. 
%for the same initial model profiles.

Equation~(11) implies that the accretion rate diverges
at $\tilde{r}=1$ (i.e., $\tau = 1$) 
and asymptotically approaches a power-law dependence
at large $\tilde{r}$ (i.e., large $\tau$), 
$\dot{\tilde{M}}_{acc} \propto \tau^{(3D_1-3)}$ if $D_1 > 2/3$, and
$\dot{\tilde{M}}_{acc} \propto \tau^{(3D_1-4/3D_1)}$ if $D_1 < 2/3$ 
(see Fig.~5). (Here again, the case $D_1 = 2/3$ is logarithmic.) 
In particular, $\dot{\tilde{M}}_{acc}$ tends to the asymptotic value 
$-\sqrt{3}.(2/\pi)$ at large $\tau$ when $D_1 = 1$ ($r^{-2}$ envelope)
(see also Appendix B). 
As was observed by H94 and FC93,  
this asymptotic limit is only achieved if the envelope 
is sufficiently large (i.e., typically $\tilde{r}_b \simgt 4$).
Otherwise the accretion rate associated with stellar formation is 
truncated abruptly as the surface of the protostellar cloud falls in. 
%Secondary 
Terminal accretion 
may continue after this stage as noted in H94 but with a character that depends 
on the density profile in the environment. We try to simulate 
this late effect here with a steep $D_2 = 0.1$ edge (see Eq.~9').\\
The accretion peak at $\tau = 1$ is due to the close approach 
of the density profile 
to being flat near $\tilde{r}=1$. Physically, if the density profile is 
strictly flat, two distinct spherical shells fall to the centre simultaneously, 
so that the frequency of shell arrival (which is the accretion rate) is infinite. 
It is in fact easy to show from the preceding theory that the 
criterion for a shell at $r$ and another at $r+\delta r$ to be accreted at 
$R=0$ at a transitory peak rate is indeed that 
$$  {d\ln{M(r)}\over d\ln{r}}\rightarrow 3. \eqno(12)$$
In fact, if the density profile is not perfectly flat 
at some point $r_N$, e.g.,  
%locally 
$\rho_o\propto r^{-2\epsilon}$ locally where $\epsilon$ is small and 
positive, then the logarithmic derivative of Eq.~(12)
attains the value $3-2\epsilon$, and one finds that the accretion rate 
for these shells will have a peak amplitude (in the usual units at $r_N$)  
$$\dot{\tilde{M}}_{acc}^{peak}=-{2\over \pi\epsilon}.\eqno(13)$$

\section{Discussion: Implications for protostellar evolution}

What we should retain from the previous analytical calculations 
is the association of 
centrally peaked, flattened, pre-stellar density profiles with 
a transitory, vigorous, post nuclear formation accretion epoch. 
We here suggest that this energetic accretion
phase coincides with Class~$0$ protostars, which would explain their 
unusually powerful jets compared to the more evolved 
Class~I YSOs (see Sect.~2.2).

%\subsection{Comparison between theory and observations} 
\subsection{Predicted versus observed accretion histories} 

%As a first step we ignore cloud-dependent effects and 
We proceed to compare the model predictions obtained in Sect.~3.3 above 
with the $\fco \, c /\lbol$ versus $\menv / \lbol^{0.6} $
diagram corresponding to the entire BATC sample. 
This diagram, which is almost 
free of any luminosity or distance effect, should  
mainly reflect the evolution of the accretion rate $\macc $.  
Indeed, as discussed by AWB93 and BATC, 
the $\menv / \lbol^{0.6} $ ratio can be used as a 
practical, quantitative indicator of embedded YSO evolution, 
which, to first order, should 
track the ratio $\menv/\ms$ of envelope to stellar 
mass and thus decrease with protostellar age.\footnote{ 
The envelope masses ($\menv$) used in the diagrams shown by BATC 
were derived from 1.3~mm maps  
by integrating the observed dust continuum emission over a region  
$\sim 1'$ in diameter (corresponding to $\sim 10^4$~AU at $d = 160$~pc). 
A mass opacity $\kappa_{1.3} = 0.01$~cm$^2\,$g$^{-1}$ was assumed (cf. AM94).} 
As to the CO outflow momentum flux $\fco $, it is related to 
the accretion rate $\macc$ by 

$$ \fco = [\fent \, (\mwin/\macc)\,\vwin] \times \macc , \eqno(14) $$
 
where $\fent$ is the entrainment efficiency, $\mwin$ is the mass-loss rate 
of the underlying driving wind or jet,  
and $\vwin$ is the wind velocity. As discussed 
in BATC, $\fent \, (\mwin/\macc)\,\vwin $ is  
unlikely to vary much during the protostellar phase, so that 
$\fco$ should reflect the variations of the accretion rate $\macc $. 
However, because as a general trend more massive cores tend to form 
more massive and more luminous YSOs associated with more 
energetic outflows (e.g. Saraceno et al. 1996), $\fco$ also depends on 
luminosity or stellar mass. For this reason, we use the quantity 
$\fco \, c /\lbol$ instead which should be less sensitive to 
%initial conditions 
the initial cloud mass/density.  
To the extent that there is no significant luminosity evolution 
for a given protostellar source (see Kenyon \& Hartmann 1995 and BATC),
$\fco \, c /\lbol$ should more 
homogeneously trace the intrinsic temporal variations of $\macc$ 
than $\fco$. 

The $\fco \, c /\lbol$ versus $\menv / \lbol^{0.6} $ diagram observed by BATC,
which is shown in Fig.~7,  
suggests that the accretion rate of Class~0 sources is a factor of $\sim 10$ 
larger on average than that of Class~I sources (see also Sect. 4.3.1 of BATC). 
By comparison with Fig.~6, it is tempting 
to identify the Class~0 phase with the short period of vigorous accretion 
predicted by the theory of Sect.~3.3 after the formation of 
the hydrostatic stellar nucleus, and the Class~I phase with 
the longer period of constant $\macc$ predicted at late times when 
$D_1 =1$ ($r^{-2}$ envelope).  
%
%  dick_fig7b.greg  -- LET MACC MACC*2e2 -- LET MENVT MENVT*0.65
%
\begin{figure}
\centerline{\hbox{\psfig{file=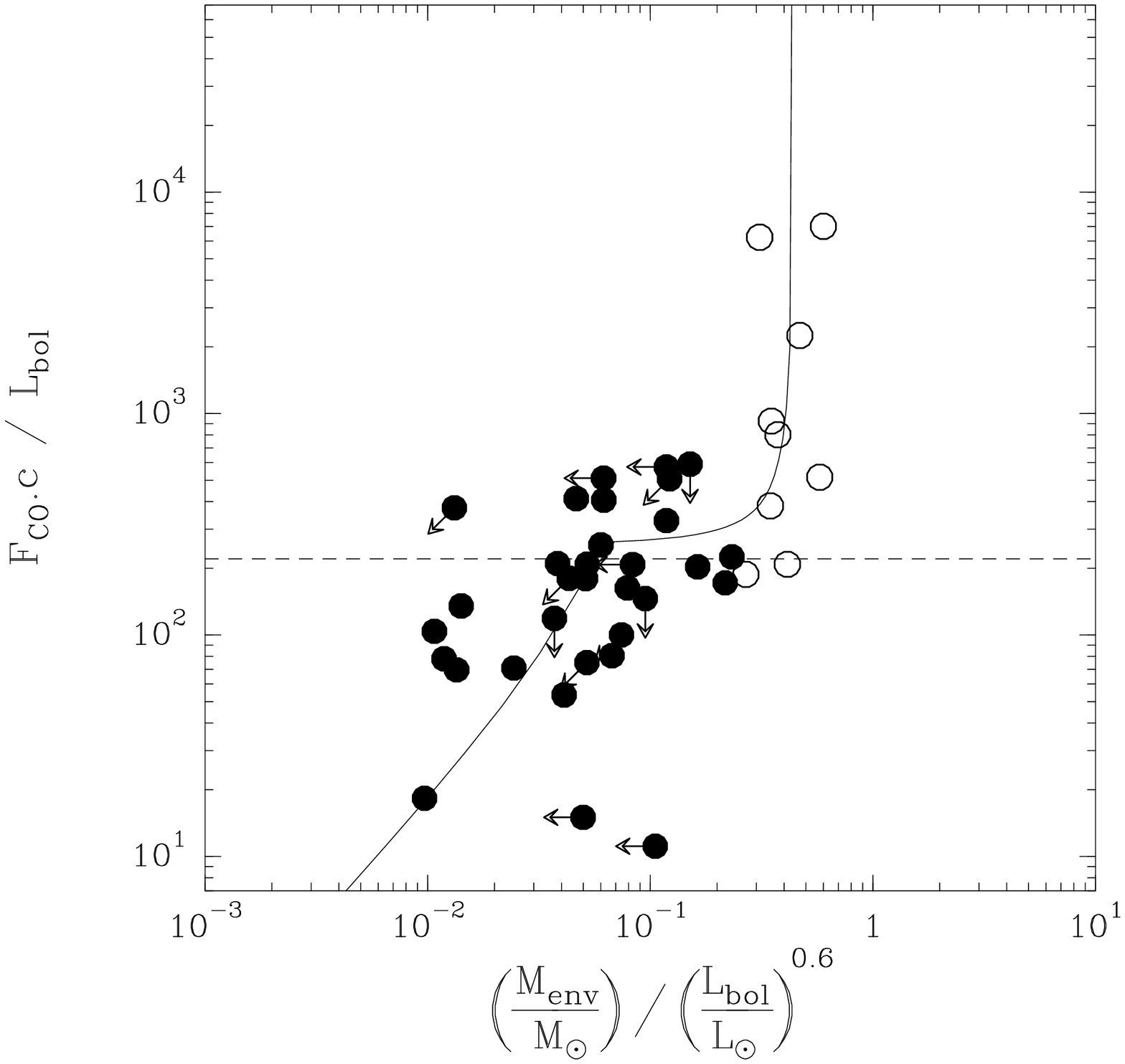,height= 8 cm,angle=270}}} 
%\vspace{8cm}
\caption{$\protect\fco\,\mbox{c}/\protect\lbol$ (dimensionless) versus
$\protect\menv/\protect\lbol^{0.6}$ ($\protect\menv$ and
$\protect\lbol$ in units of $\protect\sm$ and $\protect\sl$) for the
sample of Class~0 (open circles) and Class~I (filled circles) sources
studied by Bontemps et al. (1996). (The arrows on the data points
indicate upper limits.) $\protect\fco\,\protect\mbox{c}/\protect\lbol$
can be taken as an empirical tracer of $\protect\macc$ provided that
there is no significant luminosity evolution.
$\protect\menv/\protect\lbol^{0.6}$ is an evolutionary indicator which
decreases with protostellar age. The eye-fitted solid curve shows the
accretion rate history predicted by the model of Sect.~3.3.2 assuming
$D_1 = 1$, $D_2 = 0.1$, and $\tilde{r}_b = 1.6 $ (see text). The
dashed line shows the corresponding asymptotic value of
$\protect\fco\,\protect\mbox{c}/\protect\lbol$ for comparison with
Fig.~5 and Fig.~6.}
\end {figure}

To make quantitative comparisons, we use 
an independent observational constraint on the average values of the
accretion rate during the Class~I and Class~0 phases. 
The lifetime of Class~I sources is measured  
to be $\Delta t_I \sim 2 \times 10^5$~yr in clouds 
forming mainly low-mass stars 
such as $\rho$~Ophiuchi and Taurus
(e.g. Wilking et al. 1989, Kenyon et al. 1990, Greene et al. 1994).
The lifetime of Class~0 sources is estimated 
to be an order of magnitude shorter, $\Delta t_0 \simgt 10^4$~yr
(see AM94 and Barsony 1994).  
In this paper, we adopt the conceptual definitions of the 
Class~0 and Class~I stages introduced by AWB93 (see also 
AM94). The Class~0 stage corresponds to protostars 
which have developed a hydrostatic stellar core but still have 
$\menv > \ms $, i.e., objects which have accreted less 
than a half of the total cloud 
mass; in our model, the formation of the central hydrostatic object 
ends when the flat pre-stellar nucleus has fully collapsed, 
thus Class~0 sources 
have $M_N < \ms < M_{cloud}/2$. The Class~I stage corresponds to 
embedded sources with smaller envelope masses: $\menv < \ms $. 
We also define the boundary between Class~I and Class~II sources by 
$\ms = 0.99 \times M_{cloud}$, i.e., $\menv = 0.01 \times M_{cloud}$.
This allows for residual (disk) accretion during the (Class~II) 
pre-main sequence phase:  a $1\ \sm$ star may typically accrete up to the mass 
of the minimum solar nebula, i.e., $0.01\ \sm$, at the Class~II stage  
($0.01\ \sm$ is indeed the median circumstellar disk mass 
measured around Class~II T Tauri stars -- Beckwith et al. 1990, 
AM94). 
   
Within the framework of the model introduced in Sect.~3.3.2 the
mass function $\tilde{M}(\tilde{r})$ is easy to invert, and given 
the above definitions the 
lifetime ratio $\Delta t_I/\Delta t_0 $ can be calculated analytically 
as a function of $\tilde{M}_{cloud}$ or $\tilde{r}_b$:

$$\Delta t_I/\Delta t_0 = (\tau_I-\tau_0)/(\tau_0-1), \eqno(15) $$ 

where 
$$\tau_0 = [\tilde{r}(\tilde{M}_{cloud}/2)]^{3/2}/(\tilde{M}_{cloud}/2)^{1/2}, 
\eqno(16)$$ 
and
$$\tau_I = [\tilde{r}(0.99\tilde{M}_{cloud})]^{3/2}/(0.99\tilde{M}_{cloud})^{1/2}.
\eqno(16')$$
This model ratio is plotted in Fig.~8 as a function of the boundary radius 
$\tilde{r}_b$, for the same three values of $D_1$ as in Figs. 4, 5, 6.
%
%  timescale_fig.greg
%
\begin{figure}
\centerline{\hbox{\psfig{file=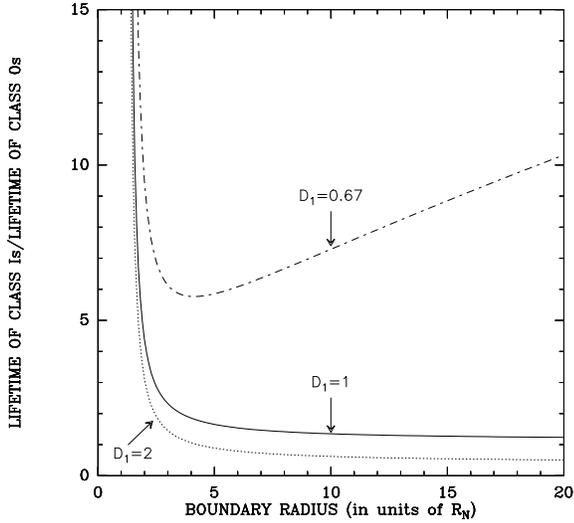,height= 8 cm,angle=270}}} 
%\vspace{8cm}
\caption { Predicted ratio of the duration of the Class~I phase $\Delta t_I$ 
to the duration of the Class~0 phase $\Delta t_0$ as a function of the 
pre-stellar boundary radius $\tilde{r}_b$ within the model of Sect.~3.3.2
(see text for formal definitions of the Class~0 and Class~I phases). 
Three representative cases are shown: $r^{-1}$ envelope ($D_1 = 2$, 
dotted curve), $r^{-2}$ envelope ($D_1 = 1$, solid curve), 
$r^{-3}$ envelope ($D_1 = 2/3$, dash-dotted curve). In all three cases,
a sharp outer boundary ($D_2 = 0.1$) has been assumed beyond $\tilde{r}_b$.
}  
\end {figure}
It can be seen from Fig.~8 that the observational constraint 
$\Delta t_I/\Delta t_0 \sim 10$ can be satisfied by the $D_1 = 1$ 
model (standard case) only if the pre-stellar boundary radius
is small ($\tilde{r}_b \sim 1.6$), corresponding to a small value of the  
normalized cloud mass ($\tilde{M}_{cloud} \sim 3$). 
Furthermore, the predicted $\Delta t_I/\Delta t_0 $ ratio decreases 
rapidly towards an asymptotic value of 1 as $\tilde{r}_b $ increases  
(one already has $\Delta t_I/\Delta t_0 \simlt 2$ 
for $\tilde{r}_b = 3.5 $, i.e., $\tilde{M}_{cloud} \sim 9$). 
The $D_1 = 2$ case is qualitatively similar.  
In contrast, for sharper initial density profiles 
such as $r^{-3}$ ($D_1 = 2/3$), the predicted 
lifetime ratio $\Delta t_I/\Delta t_0 $ 
is $\simgt 10$ for a wide range of $\tilde{r}_b $ values. 
 
Since the mean stellar mass in the BATC sample
is $<M_\star> \simlt 1\ \sm$, the average accretion rate 
during the  Class~I phase can be estimated as\\
$<\macc>_I \, \approx \, <M_\star>/(2\Delta t_I) 
\sim 2.5 \times 10^{-6}\ \myr $.
 %in the BATC sample.
In addition, one has 
$$<\macc>_0/<\macc>_I \, \approx \, \Delta t_I/\Delta t_0 \times 
(1-\frac {2}{\tilde{M}_{cloud}}). \eqno(17)$$ 

Therefore, if $\tilde{M}_{cloud}$ is sufficiently larger than 2,
the lifetime constraint $\Delta t_I/\Delta t_0 \sim 10$ implies\\ 
$<\macc>_0/<\macc>_I \, \sim 10$, 
in reasonable agreement with the drop of 
$\fco \, c /\lbol$ seen in Fig.~7 between Class~0s and Class~Is.

To reduce the number of free parameters, we will assume 
$D_1 = 1$ and $D_2 = 0.1$ from now on. With these values,
the model profile accounts for the nearly $r^{-2}$ `envelope' that  
is observed in most pre-stellar cores 
just outside the flat central region (WSHA, AWM96). It also 
features a steep 
%($r^{-20}$) 
outer edge, qualitatively similar to 
those revealed by some recent infrared absorption studies of dense cores 
(Dent et al. 1995, Abergel et al. 1996).\\
The boundary radius 
$\tilde{r}_b$ is the only 
parameter of the model that we will vary in order to match the observations.
As discussed above, the lifetime constraint suggests $\tilde{r}_b \sim 1.6$. 
In Fig.~7, we have superposed as a solid curve 
the accretion history predicted by the model when 
$\tilde{r}_b = 1.6$ (and $D_1 = 1$, $D_2 = 0.1$) to the 
$\fco \, c /\lbol$ versus $\menv / \lbol^{0.6} $ diagram 
observed by BATC. 
To scale the model along the x-axis, we adopted
$M_{cloud} = 1\ \sm$, based on the average value of $\menv$ for 
the Class~0 sources in the BATC sample. 
The outflow efficiency factor $\fent \, (\mw/\macc)\, \vwin $ was  
adjusted so as to make the model match the typical value 
$\fco\,\mbox{c}/\lbol \sim 200$ observed among the Class~I sources 
of the BATC sample. 
Given an average bolometric luminosity 
$<\lbol> \, = 2\ \sl $ (BATC), this corresponds to 
a typical momentum flux $<\fco> \, \sim 8 \times 10^{-6}\,\mkmsyr$ for 
Class~I sources.
At late times the ($D_1 = 1$) model predicts a constant accretion 
rate $M_{acc}^0$ 
corresponding to the standard Shu value (see Appendix B). 
If we adopt $M_{acc}^0 \sim 2 \times 10^{-6}\,\myr$, which is 
appropriate for a $T = 10$~K cloud and in rough agreement with the value 
estimated above from the lifetime of Class~I sources, then we find that the
%`best-fit' 
best `eye-fit'
model has an efficiency $\fent \, (\mw/\macc)\, \vwin \sim 4\,\kms$. 
We note that this is a factor of 4 lower than
the efficiency used in BATC which corresponds to the canonical values
$\vwin = 150\,\kms$, $\mw/\macc = 0.1$, and $\fent =1$. A possible explanation
for this difference is that the entrainment efficiency is actually lower than 1,
$\fent \sim 0.25$.
In spite of this discrepancy, the overall agreement between 
the shape of the predicted accretion history and the observations 
is quite encouraging given the simplicity of the present modeling.
In particular, it should be noted that the range of initial 
$M_{cloud}$ values 
spanned by the BATC sample is certainly only partly accounted for by dividing 
$\fco $ by $\lbol $ and plotting 
$\fco \, c /\lbol$ versus $\menv / \lbol^{0.6} $.\\

A potential problem of the $D_1= 1$ model shown in Fig.~7 is that 
it apparently leads to very short timescales. 
For $\tilde{r}_{b} = 1.6$, the predicted 
Class~0 and Class~I lifetimes are indeed 
%$\Delta t_0 = \tau_0 - 1 = %[\tilde{r}(\tilde{M}_{cloud}/2)]^{3/2}/(\tilde{M}_{cloud}/2)^{1/2}$. 
$\Delta t_0 \sim 0.04\, t_{ff}(N)$ and 
$\Delta t_I \sim 0.37\, t_{ff}(N)$ in units of the nuclear 
region free-fall time.  
If we take the mean density $n_{H_2} \sim 2 \times 10^5$~cm$^{-3}$ observed in
the flat central region of L1689B (AWM96) as a representative example, 
we obtain $t_{ff}(N) \sim 7 \times 10^4$~yr, and thus 
$\Delta t_0 \sim 2500$~yr and
$\Delta t_I \sim 2.6 \times 10^4$~yr, which are at least a factor of $\sim 4$ 
%almost an order of magnitude 
shorter than the observed values of $\Delta t_0 $ and $\Delta t_I $.\\  
However, in actual fact, 
the collapse is probably less violent than indicated by our
self-similar, pressure-less calculations.  
The more realistic numerical calculations of FC93 and Tomisaka (1996), 
which fully account for pressure support (including magnetic support in the 
latter case), yield qualitatively similar accretion histories but 
quantitatively longer timescales. In particular, the core formation time 
since the initiation of collapse is a factor of 
$\sim 4$--$5$ times 
larger than the free-fall time $t_{ff}(N)$ in these calculations. 
If we use this factor to roughly estimate the slowing down
influence of pressure effects, 
then we find $\Delta t_0 \sim 10^4$~yr 
and $\Delta t_I \sim 10^5$~yr, in 
better agreement with the observed lifetimes.

\subsection{Accretion luminosities of Class~0 and Class~I sources}

If, as proposed here, $\macc$ is a factor of $\sim 10$ larger 
for Class~0 sources than for Class~I sources, one may naively 
expect the former to have much higher accretion luminosities than the latter.  
This would be in apparent contradiction with observations 
which suggest that 
Class~0s are not significantly over-luminous compared to Class~Is. 
For instance, in the embedded YSO sample of BATC, 
the ratio of the average bolometric luminosities 
for the two classes was $<\lbol>_0/<\lbol>_I \sim 1.6$. (Note however that 
the typical luminosity of Class~0 sources in a given cloud is relatively
uncertain observationally due to the rarity of these objects.) 

In reality, several factors contribute to reduce the theoretical 
accretion luminosity (L$_{acc} = GM_\star \dot{M}_{acc}/R_\star$)
at the Class~0 stage, making it comparable to the
Class~I accretion luminosity:\\  
(1) The central stellar mass $M_\star$ is smaller for Class~0 sources;\\ 
(2) The stellar radius $R_\star$ is likely to be larger if $\macc$ is 
higher, since one expects the rough scaling $R_\star \propto \macc^{1/3}$
(Stahler, Shu, \& Taam 1980; Stahler 1988);\\
(3) The amount of accretion energy dissipated in the wind  
can be expected to be larger for Class~0s than for Class~Is 
(and could be a significant fraction of the total accretion energy).

Ignoring effect (3) for a moment, it is easy to compute a rough theoretical 
estimate for the ratio\\ 
$<\lacc>_0/<\lacc>_I$. Indeed,\\
$<\lacc>_0 = {1\over \Delta t_0}\, \int_1^{\tau_0}\, 
{GM_\star\dot{M}_{acc} \over R_\star}\, d\tau$, and thus
%$$<\lacc>_0= {1\over \Delta t_0}\, \int_1^{\tau_0}\, 
%{GM_\star\dot{M}_{acc} \over R_\star}\, d\tau 
$$<\lacc>_0 \approx {1\over \Delta t_0}\, {G\over <R_\star >_0}\,
\int_{M_N}^{M_{cloud}/2}\, 
M_\star\, dM_\star. \eqno(18)$$
Likewise, $<\lacc>_I= {1\over \Delta t_I} \int_{\tau_0}^{\tau_I}\, 
{GM_\star\dot{M}_{acc} \over R_\star}\, d\tau$, and
%$$<\lacc>_I= {1\over \Delta t_I} \int_{\tau_0}^{\tau_I}\, 
%{GM_\star\dot{M}_{acc} \over R_\star}\, d\tau
$$<\lacc>_I \approx {1\over \Delta t_I}\, 
{G\over <R_\star> _I}\,\int_{M_{cloud}/2}^{M_{cloud}}\, 
M_\star\, dM_\star. \eqno(18')$$
Thus,
$${<\lacc>_0 \over <\lacc>_I} \approx {1\over 3}\, 
{\Delta t_I \over \Delta t_0}\, {<R_\star> _I \over <R_\star> _0}\, 
(1-{4\over \tilde{M}_{cloud}^2}). \eqno(19)$$
In these equations, $<R_\star> _0$ and $<R_\star> _I$ are appropriately 
time-averaged values of the stellar radius at the Class~0 and Class~I 
stages, respectively. If the scaling $R_\star \propto \macc^{1/3}$ is 
valid, then one predicts 
${<\lacc>_0 \over <\lacc>_I} \simlt 1.5$ for 
${\Delta t_I \over \Delta t_0} \sim 10$, which is consistent with 
existing observations.\\
The magnitude of effect (3) is more difficult to assess since it depends 
on the still poorly understood mechanism of accretion/ejection. 
In the case of the X-celerator mechanism, the fraction of power
extracted by the wind roughly scales as the ratio $R_\star /R_X$
%$\rs/\rx$,
where 
%$\rx$ 
$R_X$ is the disk truncation radius (e.g. Shu 1995).
%[e.g. Shu 1995, RevMexAA (Ser. de Conf.), 1, 293].
As the accretion rate decreases and the centrifugal disk radius increases, 
we may expect $R_X$ to increase, thereby reducing the  
fraction of power released in the wind. The order
of magnitude of this effect may be another factor of $\sim 2$ in $\lacc$.

In conclusion, the combined effects of (1), (2), (3) above are likely  
to render the luminosities of Class~0 sources similar to those of Class~I
sources.

%\subsection{Cloud-dependent effects}
\subsection{Differences between $\rho$ Ophiuchi and Taurus}

When comparing model predictions with observations in Sect.~4.1 we considered 
all the YSOs of the BATC sample together, regardless of their parent clouds. 
However, if we consider the  
$\fco\,\mbox{c}/\lbol$ versus $\menv / \lbol^{0.6} $ diagrams 
of $\rho$~Ophiuchi (Fig.~9) and Taurus (Fig.~10) separately, clear differences 
become apparent. In the diagram of Fig.~9, 
the Class~0 sources of $\rho$~Oph (VLA~1623 and IRAS~16293)
clearly stand out as a distinct group characterized by values of 
$\fco\,\mbox{c}/\lbol$ and $\menv / \lbol^{0.6} $ which are both an order
of magnitude larger than the corresponding values for Class~I sources.  
In fact, this clear contrast observed in $\rho$~Oph between Class~0 and
Class~I objects was part of the original motivation for introducing the 
Class~0 as a new class of YSOs (see AWB93). On the other hand, no such 
contrast is observed in the Taurus diagram of Fig.~10, where there is a
much better continuity between Class~0 and Class~I sources. In other words,
the Class~0 candidates of Taurus may merely correspond to ``extreme Class~I'' 
sources.
 
%
%   dick_fig7bo.greg  -- LET MACC MACC*2.5e2 -- LET MENVT MENVT*0.65
%
\begin{figure}
\centerline{\hbox{\psfig{file=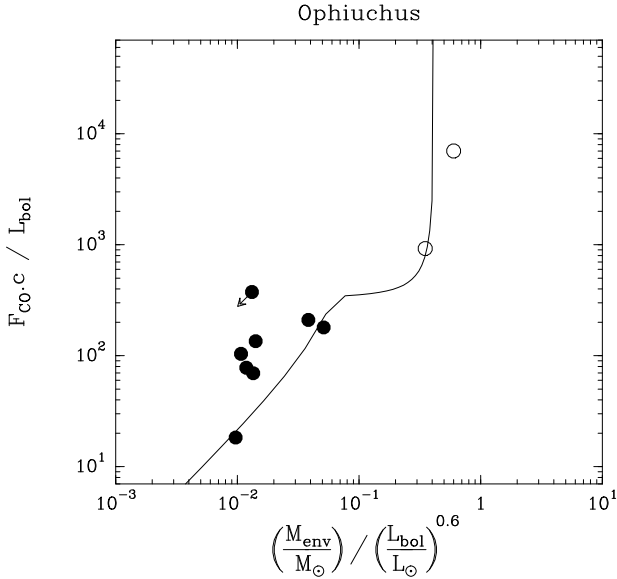,height= 8 cm,angle=270}}} 
%\vspace{8cm}
\caption { $\protect\fco\,\mbox{c}/\protect\lbol$ versus $\protect\menv/\protect\lbol^{0.6}$ diagram 
for the sub-sample of 
Class~0 (open circles) and Class~I (filled circles) sources 
observed by BATC in $\rho$ Ophiuchi. 
(The arrow indicates upper limits in both variables.)
The eye-fitted solid curve shows 
the accretion rate history predicted under the model 
of Sect.~3.3.2  
with  $\protect\tilde{r}_b = 1.5$, $D_1 = 1$, and $D_2 = 0.1$ }  
\end {figure}
%
%
%   dick_fig7bt.greg  -- LET MACC MACC*2e2 -- LET MENVT MENVT*0.65
%
\begin{figure}
\centerline{\hbox{\psfig{file=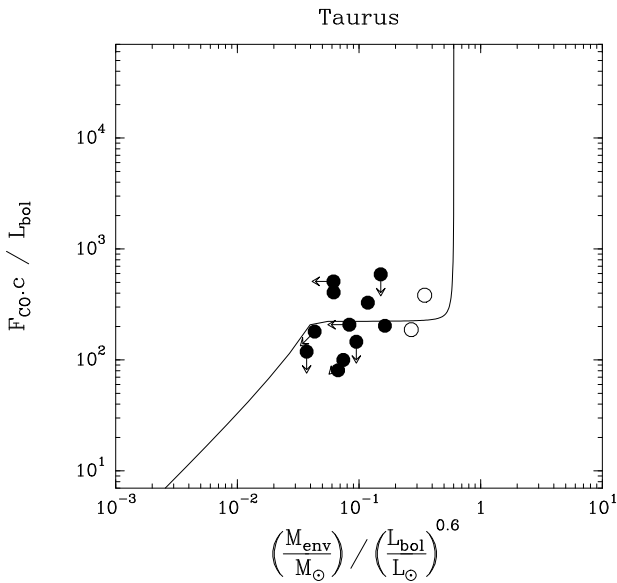,height= 8 cm,angle=270}}} 
%\vspace{8cm}
\caption { 
$\protect\fco\,\mbox{c}/\protect\lbol$ versus 
$\protect\menv/\protect\lbol^{0.6}$ diagram for the sub-sample of candidate
Class~0 (open circles) and Class~I (filled circles) protostars 
observed by BATC in Taurus.  
(The arrows on the data points indicate upper limits.)
The eye-fitted solid curve shows 
the accretion rate history predicted by the model of Sect.~3.3.2 
assuming $\tilde{r}_b = 5$, $D_1 = 1$, and $D_2 = 0.1$ }  
\end {figure}
Comparison of Fig.~9 and Fig.~10 suggests that $\rho$~Oph and Taurus YSOs 
follow different accretion histories. We suggest that this directly results  
from differences in initial conditions. 
Indeed, an important feature of the accretion model 
proposed in Sect.~3.3.2 (see also FC93) 
is that it predicts a significant drop of the mean accretion rate from the Class~0 to the Class~I stage {\it only} 
if the radius $r_N$ of the 
flat inner region in the initial profile is a large fraction of the 
cloud boundary radius $r_b$, i.e., if $\tilde{r}_b$ is small (see Fig.~8).  
Thus, the observed difference 
between Fig.~9 and Fig.~10 could be accounted for, at least qualitatively, by  
the present models if 
%$\tilde{M}_{cloud}$ 
the boundary radius $\tilde{r}_b$ were significantly smaller in 
$\rho$~Ophiuchi than in Taurus.

Interestingly enough, such a trend in the values of 
%$\tilde{M}_{cloud}$ 
$\tilde{r}_b$ seems to be borne out by independent 
(sub)millimeter continuum results on the 
structure of protostellar envelopes and dense cores. 
In Taurus where YSOs 
form in relative isolation, protostellar envelopes are observed to be extended
over radii $\simgt 10^4$~AU 
%with estimated density gradients consistent with 
%$\rho(r) \propto r^{-1.5}$ or $\rho(r) \propto r^{-2}$ 
(e.g. Motte et al. 1996). 
In contrast, $\rho$~Ophiuchi is a star-forming cluster 
where the fragmentation size scale is significantly smaller than $10^4$~AU
(e.g. Motte et al. 1997), 
so that the radius of the `sphere of influence' of a given YSO must be 
$r_b < 4000$~AU. 
The density profiles of pre-stellar cores are always found 
to flatten out at radii $r < r_N \simlt 4000$~AU (see Sect.~2.1 and 
references therein), with no clear difference between $\rho$~Oph and Taurus. 
These observations suggest that, if bounded at all, 
the Taurus cores have a large boundary radius $\tilde{r}_b > 3$,  
%corresponding to $\tilde{M}_{cloud} \simgt 15$, 
while the $\rho$~Oph cores 
have a smaller boundary radius $\tilde{r}_b \simgt 1$.
%corresponding to $\tilde{M}_{cloud} \simgt 3$.   

Models constructed with $\tilde{r}_b = 1.5$ and $\tilde{r}_b = 5$, 
corresponding to $\tilde{M}_{cloud} \approx 2.8 $ and 
$\tilde{M}_{cloud} \approx 13.9 $, 
have been superposed on the diagrams of Fig.~9 and Fig.~10, respectively.   
The models shown in the figures  
predict $<\macc>_0/<\macc>_I \sim 4$ in Ophiuchus and\\
$<\macc>_0/<\macc>_I \sim 1.4$ in Taurus, in reasonable agreement with observations.
As in Sect.~4.1, the outflow efficiency factor $\fent \, (\mw/\macc)\, \vwin $ was adjusted in order to `fit' the observations (by eye).
We adopted standard, asymptotic 
accretion rate values of $M_{acc}^0 \sim 10^{-5}\,\myr$ in Ophiuchus and 
$M_{acc}^0 \sim 2 \times 10^{-6}\,\myr$ in Taurus 
(e.g. Adams, Lada, \& Shu 1987), and we used 
average bolometric luminosities of $<\lbol> \, = 6\, \sl $ 
and $<\lbol> \, = 0.7\, \sl $ 
for the Class~I sources of the two clouds, respectively (see BATC). 
This procedure yields an outflow efficiency 
$\fent \, (\mw/\macc)\, \vwin \sim 6\,\kms$ in 
$\rho$~Oph, and $\fent \, (\mw/\macc)\, \vwin \sim 1.4\,\kms$ in Taurus. 
(Taken at face value, these numbers point to a somewhat more efficient
ejection process in $\rho$~Ophiuchi than in Taurus.)\\

In $\rho$~Oph, the model fit of Fig.~9 suggests that the Class~0 
sources are in the initial phase of vigorous accretion 
while the Class~I sources may be  
in the terminal accretion phase. 
The latter is not surprising since, due to the small fragmentation size scale of
this cloud, the expansion wave characterizing all collapse solutions at $t > 0$ 
(Shu77, WS85) will reach  the boundary of any given protostellar envelope/core  
in a time $t_b = r_b/a_s \sim 5 \times 10^4$~yr (assuming $a_s = 0.35\, \kms$),
shorter than the typical Class~I lifetime.\\
In contrast, all the Taurus sources of Fig.~10 appear to be
%is consistent with all the embedded sources observed in Taurus being 
in the `asymptotic' phase 
during which the accretion rate is approximately 
that predicted by the Shu theory.\\ 
The phase of enhanced accretion is thus apparently observable in $\rho$~Oph but 
not in Taurus. We suggest this is related 
to the fact that core collapse is probably induced by the
impact of a weak interstellar shock wave in the first case (e.g. Vrba 1977,
Loren \& Wootten 1986) and self-initiated by ambipolar diffusion in 
the second case (e.g. Lizano \& Shu 1989, Mouschovias 1991).\\
%accreting approximately at the rate predicted by the Shu theory.\\ 
We conclude that the standard theory of Shu and co-workers is roughly
adequate in Taurus but incomplete in $\rho$ Oph. In the latter cloud, 
the accretion scenario advocated here provides a better fit to the data.  
Further support to this view comes from the fact that the only 
candidate `isothermal protostars' known to date have precisely been found in regions of multiple star formation like $\rho$ Oph, 
in the form of very dense compact clumps only seen at 
submillimeter wavelengths (Mezger et al. 1992a,b; Chini et al. 1997).
Indeed, while the initial isothermal collapse 
phase is vanishingly short in the standard Shu picture (see Sect.~1), it 
should last for a significant fraction of the total collapse/accretion time
if $r_N/r_b$ is large and if the scenario proposed in this paper 
%described in the previous section 
is approximately correct (see, e.g., Sect. 3.3.1). 

\section{Summary and conclusions}
 
The main points of our paper may be summarized as follows: 

\begin{enumerate}
 \item The radial density gradient of pre-stellar cores is
typically flatter than $\rho(r) \propto r^{-1}$ near their centers and 
approach $\rho(r) \propto r^{-2}$ only beyond a few thousand AU (WSHA, AWM96,
Fig.~1).
In some cases, sharp outer boundaries, much steeper than 
$\rho(r) \propto r^{-2}$, are observed at a finite core radius 
(e.g. Abergel et al. 1996). This raises the possibility that, 
in some instances at least, 
the initial conditions for protostellar collapse 
depart significantly from a singular isothermal sphere. 
 \item The outflow momentum flux of embedded protostellar sources
correlates very well with their circumstellar envelope mass
(see Fig.~2 and Fig.~7). This suggests 
that the mass ejection and mass accretion rates of protostars 
both decline 
%more or less continuously 
with time during protostellar evolution from Class~0 to Class~I sources (BATC). 
 \item Recent self-consistent hydrodynamical
%numerical 
calculations of protostellar collapse indicate that initial
conditions characterized by centrally flattened density profiles 
%If the initial density profile is centrally flattened, 
%gravitational collapse will 
lead to a transitory phase of enhanced accretion immediately 
following the formation of the central hydrostatic protostar. 
This behavior is found regardless of whether the influence of 
magnetic fields is ignored (FC93) or fully 
accounted for (Tomisaka 1996). The nature of the transitory accretion peak is,
however, left somewhat unclear by these numerical simulations.
 \item On the basis of (3), we propose that (2) is a direct consequence of (1).
 \item In order to elucidate the physical origin of (3), we analytically follow 
the history of protostellar accretion, using  
Lagrangian calculations based on a 
new type of self-similar `gravity-dominated' collapse solutions   
(Sect.~3.2, Sect.~3.3, and Appendix~A).
We claim that these new solutions provide a better description of the collapse
than the usual `pressure-dominated' self-similar solutions (e.g. Shu77, WS85), 
when pressure becomes negligible in the self-gravitating flow sometime after 
the onset of collapse, prior to stellar core formation. 
Comparison with the numerical simulations shows that the gravity-dominated solutions are adequate, at least qualitatively, in the 
supersonic region.
%These solutions are determined by the shape of the density (or mass) profile 
%at some epoch in the region of supersonic flow.
%
 \item In agreement with (1), we start our supersonic calculations at 
$t = t_o < 0$ from an idealized pre-stellar core consisting of 
a strictly flat inner plateau up to a radius $r_N$, an $r^{-2}$ `envelope' up to
$r_b$, and a steeper power-law `environment' farther out (Fig.~3).  
In our gravity-dominated description, the central plateau region 
first collapses homologously to form a {\it finite-mass} 
hydrostatic stellar core at $t = 0$.
Observationally, this initial phase, which does not exist in the 
standard Shu picture, should correspond to `isothermal protostars', i.e., 
supersonically collapsing cloud fragments with no central YSOs.  
It is followed at $t > 0$ by the main accretion/ejection phase, during which
the non-zero central point mass accretes the surrounding envelope. 
As long as the gravitational influence of the initial point mass 
is significant, the accretion rate remains higher than the Shu value, 
$a_{eff}^3/G$.
%($\dot{M}_{acc} = a_{eff}^3/G$).
It then quickly converges towards $a_{eff}^3/G$.
%the Shu value. 
At late times, accretion of the outer environment  
leads to a terminal phase of residual 
accretion/ejection, during which the accretion rate declines below the Shu 
value (see Fig.~5).
 \item We use our analytical model to fit
the diagram of outflow efficiency $\fco \, c /\lbol$ 
versus normalized envelope mass $\menv / \lbol^{0.6} $
obtained by BATC (see Fig.~7).
%which indicates a clear decline of outflow/inflow
%strength from Class~0 to Class~I sources. 
%in star-forming clusters.  
To the extent that there
is a direct proportionality between accretion and ejection, this diagram 
should provide an empirical measure of the accretion history of the
sampled protostellar objects. 
A good overall fit is found when the
model boundary radius  $r_b$, 
%defined as the outer radius of the envelope or 
%the inner radius of the environment, 
is not much larger than the radius 
$r_N$ of the flat inner plateau, i.e., $r_b \sim 1.6\, r_N$. This requires that 
the fraction of cloud mass in the central plateau region be 
relatively large, $M_N/M_{cloud} \sim 30$~\%. (However, since the collapse 
is in fact less violent than in our pressure-less calculations, 
these values of $r_b$ and $M_N$ 
should be taken as indicative, and are likely to represent only
a lower limit and an upper limit, respectively.)
 \item Based on (7), we tentatively associate the short 
%initial 
period of energetic accretion/ejection predicted by our model 
at the beginning of the accretion phase 
%(see also FC93 and Tomisaka 1996) 
with the observationally-defined Class~0 stage (Sect.~4.1 and Fig.~7).  
In this view, Class~I objects are more evolved and correspond to 
the longer period of moderate accretion/ejection when the accretion rate 
approaches the Shu value. (See, however, point 10 below.)
 \item We also find 
that the Ophiuchus $\fco \, c /\lbol$ versus $\menv / \lbol^{0.6} $ diagram
differs markedly from the corresponding diagram in Taurus: 
%Taurus $\fco \, c /\lbol$ versus $\menv / \lbol^{0.6} $ diagram: 
a clear contrast 
between Class~0 and Class~I objects is observed in Ophiuchus (Fig.~9)
which is not seen in Taurus (Fig.~10). 
This points to different accretion histories 
%suggests that the YSOs of 
in these two nearby star-forming clouds, which  
we interpret as arising from differences in initial conditions (Sect.~4.3).  
Both outflow and dense core observations suggest that 
the relative importance of the central plateau region in the initial density profile
is significantly larger in Ophiuchus 
($r_N \sim r_b $, $M_N/M_{cloud} \simgt 30$~\% according to our approximate
model) than in Taurus 
($r_N \sim 0.2\, r_b$, $M_N/M_{cloud} < 10$~\%).  
\item  According to our model fit of the Ophiuchus 
$\fco \, c /\lbol$ versus $\menv / \lbol^{0.6} $ diagram,
%suggests that 
most of the $\rho$~Oph Class~I YSOs  should be in their terminal 
accretion phase (see Fig.~9). 
This is not surprising since the relatively small fragmentation
length scale observed in $\rho$~Oph 
%(e.g. Motte et al. 1997) 
implies that only 
%a finite-size sphere of influence and 
a finite reservoir of mass is effectively available 
for the formation of any given protostar.
\item 
%Thus, in regions of isolated star formation such as Taurus, 
In conclusion, the `standard' theory of Shu and co-workers appears to 
describe protostellar evolution quite satisfactorily 
in regions of isolated star formation like Taurus. 
In our view, this is because in these regions most stars probably form   
following the {\it self-initiated} contraction/collapse of 
dense cores due to ambipolar diffusion.\\ 
%
%\item 
However, in star-forming clusters such as $\rho$~Ophiuchi,  
the standard Shu theory is less appopriate 
%to describe protostellar evolution. 
since it does not account for fragmentation and multiple star formation. 
%while observations indicate that the fragmentation length scale is 
%significantly less than 0.1~pc (e.g. Motte et al. 1997). 
%In this case, 
In these regions, 
star formation may be {\it induced} by the impact of (slow) shock waves 
%(e.g. Loren \& Wootten 1986), 
(e.g. Boss 1995), and protostellar cores may form  
%approach the singular $\rho \propto r^{-2}$ density distribution by 
by supersonic (or superalfv\'enic) implosion of dense clumps, 
rather than slow (subsonic) ambipolar evolution. 
In this case, the collapse/accretion history 
advocated in the present paper (Sect.~3.3) and 
summarized in (6) is likely to provide a better description (see Fig.~9 and 
Sect.~4.3).

\end{enumerate}

\acknowledgements{ We acknowledge useful discussions with Charlie Lada, 
Phil Myers, Francesco Palla, and Derek Ward-Thompson. We also thank 
Fr\'ed\'erique Motte for helping prepare Fig.~1 and Thierry Foglizzo for 
pointing out the related work by Tomisaka (1996). We are grateful to 
our referee Frank Shu for constructive criticism.}

%\newpage
\vspace{0.5cm}

{\bf Appendix A: Lagrangian collapse and self-similarity}
\vspace{0.2cm}

In Henriksen (1989 -- H89) it was shown in effect that in terms of a Lagrangian 
label $r$ defined as the radial position of an isothermal spherical shell 
at a fiducial instant $t=  t_o$ {\it when the sphere is at rest}, and 
a  variable 
$$\xi\equiv \sqrt{2GM(r)/r^3}\,(t- t_o)\eqno(A1)$$ 
($M(r)$ is the mass initially 
inside the radius $r$), the subsequent position of the shell $R(t,r)$ 
can be written as $$ R=r{\cal S}(\xi,r). $$
The `scale factor' ${\cal S}$ obeys the equation 
$$ (\partial_\xi{\cal S})^2 -1/{\cal S} + \alpha_s(r)^2 {\cal H}(\xi,r) =-1.
\eqno(A2)$$

Here, we have defined    
$\alpha_s^2\equiv a_s^2 r/GM(r)$, where $a_s$ is the sound speed, and 
${\cal H}\equiv \ln{\rho(\xi,r)/\rho_o(r)}$. The function $\rho_o(r)$ is the initial density (at $t= t_o$) 
that is related to $M(r)$ through $\rho_o=({1\over 4\pi r^2}){dM\over 
dr}$. We shall discuss the possible forms of this function further below.

In order to obtain a complete description of collapse from rest, we must add 
the equation for the density in the form 

$$e^{{\cal H}(\xi,r)}={1\over {\cal S}^3\left(1+{\partial\ln{\cal S}\over\partial\ln r} 
-{1 \over D(r)}{\partial\ln{\cal S}\over \partial\ln\xi}\right)},\eqno(A3)$$

where the function ${- 1 \over D(r)}\equiv {\partial\ln\xi\over\partial\ln r}=
{3\over 2}\left({1\over 3}{d\ln M(r)\over d\ln r}-1\right)$.  

These equations are equivalent to the equations used previously in Eulerian 
form by most other authors (e.g. FC93). 
Clearly ${\cal S}(\xi=0,r)=1$ is one boundary condition and ${\cal H}(\xi=0,r)=0$
is another, 
but in addition we must impose an equilibrium condition if this is to be a 
`natural' starting point. The equilibrium condition consists in requiring 
$\rho_o(r)$ to satisfy the Lane-Emden equation for an isothermal gas sphere. 
Consequently it is either the general function $\rho_o/\rho_o(0)\equiv 
e^{-\psi(r)}$ discussed for example in 
Chandrasekhar (1939) and used in the Bonnor-Ebert discussion of 
instability (e.g. Bonnor 1956), or 
it is $\rho_o=\lambda/r^2$ everywhere, which is the singular isothermal 
sphere (SIS). In the latter event $\alpha_s ^2 = 1/2$ is constant and  
 
$$\lambda\equiv \alpha_s^{-2} {a_s^2\over 4\pi G}. \eqno(A4) $$

Moreover the variable $\xi$ is recognizable in this case as  
 
$$\xi = \sqrt{2/ \alpha_s^2}~ {a_s (t- t_o)\over r}. \eqno(A5)$$

The SIS is exceptional in that Eqs.~(A2) and (A3) show that 
the development can be immediately  
`self-similar'; that is  ${\cal S}(\xi,r)=S(\xi)$, and  
${\cal H}=H(\xi)$ since $D(r)=1$ for all $r$. 
This is the initial condition and collapse 
first studied by Shu77 and subsequently developed by Shu and co-workers 
into the `standard model' of protostar formation. 

However, a closer inspection of Eqs. (A2) and (A3) 
combined with our knowledge of the expected parametric
collapse instability discovered by Bonnor and Ebert 
reveals that there are several other possible lines of development. These are 
all based on the assumption that the initial equilibrium is 
{\it not} that of the 
SIS, but rather a classical Bonnor-Ebert solution 
%the classical non-singular solution 
with a flattened core 
($\rho_o(0)$ is finite and $\psi\approx x^2/6$ for small 
$x\equiv \sqrt{4\pi G \rho_o(0)/a_s^2}r$) and a $r^{-2}$ halo. 

In general, if we ignore the somewhat artificial means of 
launching the collapse 
by a sudden cooling of the gas so that $a_s^2\rightarrow 0$, the development 
from the Bonnor-Ebert 
sphere is clearly {\it not} 
self-similar since $\alpha_s$ and $D$ are complicated 
functions or $r$. 
But our fundamental theoretical remark in this paper is 
that a self-similarity in terms of the general variable $\xi(t,r)$ 
(as in (A1))  can 
develop dynamically if $\alpha_s\rightarrow 0$  as the 
collapse develops. For then ${\cal S}\rightarrow S(\xi)$ by Eq.~(A2) 
and the density tends to 
%$\rho/\rho_o=S^{-3}/(1+D(r)\partial\ln S/\partial\ln \xi)$ 
$\rho/\rho_o=S^{-3}/(1-{1 \over D(r)}d\ln S/d\ln \xi)$
as used in the text. 

To place this idea in context it is worth remarking that none of the entire 
`hyperbolic' (that is characterized by a wave propagation speed) family of 
self-similar solutions found by WS85 can begin exactly from the 
Bonnor-Ebert initial 
condition either. They must then arise dynamically in the same way that we suggest 
occurs for the zero pressure flow.  The `band zero' solutions may represent 
an intermediate 
stage however because of their flattened, dis-equilibrium, central peaks, 
as discussed in the text. Such solutions are all constrained to pass through 
an instant where $\rho\propto r^{-2}$ because of their assumption of the 
strict self-similar symmetry (see e.g. Carter \& Henriksen 1991) expressed 
in Eq.~(A5) (normally authors set $ t_o=0$). This instant occurs at 
$\xi=0$ (see e.g. Eq.~1 of Hunter 1977) which is normally taken to 
coincide with $t=0$. This family of solutions may be discussed 
elegantly in terms of the Lagrangian formulation as discussed in H89. 
However since these solutions are well known (WS85), 
we pursue the novel idea in this paper as noted 
above, that the self-similar symmetry itself may change dynamically from 
that of (A5) to that determined by $\xi$ where a more general mass or density 
profile is allowed. Essentially we suppose that the symmetry changes from 
being `hyperbolic' to `elliptic', which is the symmetry appropriate to 
the local acceleration  that is present in a spherical flow dominated by gravity.

This requires that the flow evolve away from the initial equilibrium under 
the influence of a Bonnor-Ebert-Tomisaka type instability towards a density 
profile where $\alpha_s$ is small compared to ${\cal S}^{-1}-1$ 
(Eq.~A2), 
at least in some finite region.  At this instant, which we designate as 
$-t_{ff}(N)$ in the text, where the gas will no longer 
be in equilibrium but rather in supersonic flow ($(\partial S/\partial\xi)^2>>
\alpha_s^2$), the current position of the mass shells becomes the new 
comoving variable $r$. Moreover ${\cal S}\rightarrow S(\xi)$ becomes determined 
by the pressure-free version of Eq.~(1) in the text, for which we take 
the bound solution that becomes Eq.~(4) of the text near $\xi=\pi/2$.  
Here we distinguish between $ t_o$ and $-t_{ff}(N)$ to allow for an uncertain
duration between the actual initiation of collapse and the onset 
of supersonic flow
(in the text the formulae are simplified by taking $ t_o= -t_{ff}(N)$).

This scenario will be realized in the vicinity of $r_N$ for example if at 
$-t_{ff}(N)$ the density 
profile is roughly as in Fig.~3 of the text, {\it and} if $a_s^2r_N/GM_N<<1$.
We assume that this condition is attained during the evolution between 
$t= t_o$ and $t=-t_{ff}(N)$. 
From this point on we describe the flow as possessing the elliptic 
self-similar symmetry of zero-pressure self-gravitating flow.  

One may note that the L/P solution, which seems to be a better (although not 
perfect) fit to the simulation results than is the S solution (FC93),  
also develops with $\alpha_s$ small. 
In fact such a band~0 solution (WS85) is described by 
Eqs.~(A2) (with $k=0$), (A3) and (A4), and so we may calculate 
the radial velocity near the singularity as   
    
 $$v\equiv\partial_t R=dS/d\xi ~\sqrt{2GM(r)/r}=-\sqrt{8\pi G\lambda}
\equiv -a_s~\sqrt{2\over \alpha_s^2}.$$    
We have used the fact that $dS/d\xi=1$ for this solution at $t=0$ where 
$S=1$ and $H=0$ by Eq.~(A2).

The parameter $\alpha_s^2$ corresponds to the parameter $1/u_\infty$ of 
WS85. According to Larson (1969), 
%the numerical calculations of FC93, 
the above velocity should
be $\approx 3.3 a_s$, so that $\alpha_s^2\approx 0.18$ in the L/P solution.
 
However, the main message of this appendix is really 
the passage to the elliptic zero pressure self-similarity at $t=-t_{ff}(N)$,
that we are advocating based on observational arguments and the theoretical 
arguments outlined above.

%
%\newpage
\vspace{0.7cm}

{\bf Appendix B: Asymptotic value of the accretion rate}
\vspace{0.1cm}

We may make contact with the accretion 
rates found by Shu77 and by FC93 by modifying slightly 
the argument of H94. If as above $r_N$ is the initial radius 
at which the profile becomes sufficiently flat to initiate the homologous 
nuclear collapse then 
$$M_N={4\pi\over 3}\rho_N r_N^3.$$
Moreover we define here the equivalent of the `virial' radius used in 
H94 by referring to the dimensionless variable 
$\zeta\equiv r\sqrt{4\pi G\rho_c}/a_s$. 
This gives in the present context 
$$ r_N=\zeta_N {a_s\over \sqrt{4\pi G \rho_N}}.\eqno(B1)$$
%Now
If the initial sphere were exactly Bonnor-Ebert, then the difference of 
density from $\rho_N$ at $\zeta_N$ is $\propto \zeta_N^2/6$ 
(e.g. Chandrasekhar 1939).  
Thus $\zeta_N\le 1$ to satisfy the flatness criterion. However 
we have detected  
a tendency for the density profile to flatten on the outside  
so it is possible that $\zeta\ge 1$ at the initiation 
of homologous collapse. Thus we retain $\zeta_N$ here as a parameter. 
As remarked in 
H94, the preceding two relations can be rearranged into the useful forms 
$$ r_N=({3\over \zeta_N^2})~{GM_N\over a_s^2},\eqno(B2)$$ 

and 
$$ \rho_N={\zeta_N^2\over 4\pi r_N^2}~{a_s^2\over G}.\eqno(B3)$$
The units in terms of which we have calculated our accretion rate now follow as 
$$\dot{M}_N\equiv 4\pi r_N^2\rho_N \sqrt{2GM_N/r_N}= \sqrt{2\over 3}\zeta_N^3~{a_s^3 
\over G}.\eqno(B4)$$
Since moreover the asymptotic value of the accretion rate in these units was 
found above to be $-\sqrt{3}~(2/\pi)$, we predict for the accretion plateau 
that is established as the $r^{-2}$ outer region falls in the rate
$$ \dot{M}_{acc}^{asymtote}= 
- {2\sqrt 2\over \pi}\zeta_N^3~{a_s^3\over G}.\eqno(B5)$$ 
The numerical factor is $0.900~\zeta_N^3$ so that $\zeta_N=1.027$ in order 
to yield the asymptotic accretion rate predicted by the Shu solution 
(a numerical factor of $0.975$ in Eq.~B5). This value of $\zeta_N$ seems 
quite possible in view of the expected flattening.

%\newpage


\begin{thebibliography}{99}
\bibitem[1996]{Abergel}
Abergel, A., Bernard, J.P., Boulanger, F. et al. 1996, A\&A, 315, L329
%ISO special issue
%\bibitem[1991]{Abramowitz}
%Abramowitz, M., Stegun, I.A. 1970, Handbook of Mathematical Functions 
%(New York: Dover) 
%\bibitem[1991]{Adams91}
%Adams, F.C. 1991, ApJ 382, 544
\bibitem[1987]{ALS87}
Adams, F.C., Lada, C.J., Shu, F.H. 1987, ApJ 312, 788
%\bibitem[1995]{a95}
%Andr\'e, P. 1995, Ap\&SS 224, 29
\bibitem[1994]{a94}
Andr\'e, P. 1994, in ``The Cold Universe'', Ed. T. Montmerle, C.J. Lada, 
I.F. Mirabel, J. Tr\^an Thanh V\^an (Gif-sur-Yvette: Editions Fronti\`eres), 
p.~179
\bibitem[1994]{am94}
Andr\'e, P., Montmerle, T. 1994, ApJ 420,837 -- AM94
\bibitem{b4}
Andr\'e P., Ward-Thompson D., Barsony M., 1993, ApJ, 406, 122 -- AWB93
\bibitem{awm}
Andr\'e P., Ward-Thompson D., Motte, F. 1996, A\&A, 314, 625 -- AWM96
%
\bibitem[1996]{bac96}
Bachiller, R. 1996, A.R.A.A., 34, 111
\bibitem[1982]{Baren82}
Barenblatt, G. E., \& Zel`dovich, Ya. B. 1982, Ann. Rev. Fluid Mech., 125, 137
%
\bibitem{barsony94}
Barsony M. 1994, in ``Clouds, Cores, and Low-mass Stars'', Ed. D.P. Clemens \&
R. Barvainis, A.S.P. Conf. Series, 65, 197
%
\bibitem[1995]{basu95}
Basu, S.,  Mouschovias, T.Ch. 1995, ApJ, 453, 271
%
\bibitem[1990]{beckwith}
Beckwith, S.V.W., Sargent, A.I., Chini, R.S., G\"usten, R. 1990,
AJ, 99, 924 
%
%\bibitem{b6}
%Beichman C. A., Myers P. C., Emerson J. P., Harris S., Mathieu R., 
%Benson P. J.,Jennings R. E., 1986, ApJ, 307, 337
\bibitem{b7}
Benson P. J., Myers P. C., 1989, ApJS, 71, 89 -- BM89
%\bibitem[1992]{bertoldi}
%Bertoldi, F., \& McKee, C.F. 1992, ApJ, 395, 140
%\bibitem[1993]{blitz}
%Blitz, L. 1993, Protostars \& Planets III, p.~125
\bibitem[1988]{Blottiau}
Blottiau, P., Bouquet, S., \& Chi\`eze, J.P. 1988, A\&A, 207, 24
\bibitem[1956]{Bonnor}
Bonnor, W.B. 1956, MNRAS, 116, 351
%\bibitem{Bontemps} Bontemps, S. 1996, PhD thesis, University of Paris XI
%
%\bibitem{BAW} Bontemps, S., Andr\'e, P., \& Ward-Thompson, D. 1995, 
%A\&A, 297, 98
\bibitem{BATC} Bontemps, S., Andr\'e, P., Terebey, S. \& Cabrit, S. 1996, 
A\&A, 311, 858 -- BATC
\bibitem[1995]{boss95}
Boss, A.P. 1995, ApJ 439, 224
\bibitem[1995]{boss_yorke95}
Boss, A.P., \& Yorke, H.W. 1995, ApJ 439,L55
\bibitem[1991]{CH91}
Carter, B. \& Henriksen, R.N. 1991, J. Math. Phys. 32, 2580
\bibitem[1939]{Chandrasekhar}
Chandrasekhar, S. 1939, An Introduction to the Study of Stellar Structure, 
%Dover Publications, University of Chicago Press, p.155
University of Chicago Press, Chapter IV
%\bibitem[1987]{Chieze}
%Chi\`eze, J.-P. 1987, A\&A 171, 225
%\bibitem[1987]{ChiezePdF}
%Chi\`eze, J.-P., \& Pineau des For\^ets, G. 1987, A\&A 183, 98
%\bibitem[1993]{Chini93} 
%Chini, R., Kr\"ugel, E., Haslam, C.G.T., Kreysa, E., Lemke, 
%R., Reipurth, B., Sievers, A., \& Ward-Thompson, D. 1993. A\&A,
%{\bf 272}, L5
\bibitem[1997]{Chini97} 
Chini, R., Reipurth, B., Ward-Thompson, D., Bally, J., Nyman, L.A.,
Sievers, A., \& Billawala, Y. 1997, ApJ, 474, L135
%\bibitem[1995]{Choi95}
%Choi, M., Evans II, N.J., Gregersen, E.M., \& Wang, Y. 1995. ApJ, 448, 742
%\bibitem[1996]{C96}
%Ciolek, G.E. 1996, private communication
\bibitem[1994]{CM94}
Ciolek, G.E., \& Mouschovias, T.Ch. 1994, ApJ, 425, 142
%\bibitem[1995]{CM95}
%Ciolek, G.E., \& Mouschovias, T.Ch. 1995, ApJ, 454, 194
%\bibitem[1994]{crutcher94}
%Crutcher, R.M., Mouschovias, T.Ch., Troland, T.H., \& Ciolek, G.E. 1994, ApJ, 
%427, 839
\bibitem[1996]{Crutcher96}
Crutcher, R.M., Troland, T.H., Lazareff, B., \& Kaz\`es, I. 1996, ApJ, 
456, 217
\bibitem[1995]{Dent95}
Dent, W.R.F., Matthews, H.E., \& Walther, D.M. 1995, MNRAS, 277, 193
%\bibitem[1979]{emerson87}
%Emerson, D.T., Klein, U., \& Haslam, C.G.T. 1979, A\&A, 76, 92
%\bibitem[1985]{FP85}
%Falgarone, E., \& Puget, J.-L. 1985, A\&A, 142, 157 
%\bibitem[1986]{FP86}
%Falgarone, E., \& Puget, J.-L. 1986, A\&A, 162, 235
%\bibitem[1992]{FPP92}
%Falgarone, E., Puget, J.-L., \& P\'erault, M. 1992, A\&A, 257, 715
\bibitem[1995]{ferreira95}
Ferreira, J., Pelletier, G. 1995, A\&A 295,807
\bibitem[1996]{FH96a}
Fiege, J., Henriksen, R.N. 1996a, MNRAS, 281, 1038
\bibitem[1996]{FH96b}
Fiege, J., Henriksen, R.N. 1996b, MNRAS, 281, 1055
%\bibitem[1994]{foghenrik}
%Foglizzo, T., \& Henriksen, R.N. 1993, Physical Review D, 48, 4645
\bibitem[1993]{fost:chev93}
Foster, P.N., Chevalier, R.A. 1993, ApJ 416,303 -- FC93
%\bibitem[1993]{Galli}
%Galli, D., \& Shu, F.H. 1993, ApJ, 417, 220
\bibitem[1994]{Goodman}
Goodman, A.A., \& Heiles, C. 1994, ApJ, 424, 208
\bibitem[1994]{gwa94}
Greene T.P., Wilking B.A., Andr\'e P., Young E.T., Lada C.J. 1994, 
ApJ 434, 614
%\bibitem[1993]{griffin93}
%Griffin, M.J., \& Orton, G.S. 1993, Icarus, 105, 337
%\bibitem[1995]{henning}
%Henning, Th., Michel, B., \& Stognienko, R. 1995, Planet. Space Sci.
%(Special issue: Dust, molecules and backgrounds), 43, 1333
\bibitem[1989]{henrik89}
Henriksen, R.N. 1989, MNRAS, 240, 917 -- H89
\bibitem[1994]{henrik94}
Henriksen, R.N. 1994, in: Montmerle T., Lada C.J., Mirabel I.F., 
Tr\^an Thanh V\^an J. (eds.)
The Cold Universe. Editions Fronti\`eres, p.241  -- H94
\bibitem[1994]{henrikvalls}
Henriksen, R.N., Valls-Gabaud, D. 1994, MNRAS, 266, 681
\bibitem[1962]{hunter62}
Hunter, C. 1962, ApJ, 136, 594
\bibitem[1977]{hunter77}
Hunter, C. 1977, ApJ, 218, 834
\bibitem[1995]{kenyon95}
Kenyon S.J., Hartmann L.W. 1995, ApJS 101, 117
\bibitem[1990]{kenyon90}
Kenyon S.J., Hartmann L.W., Strom K.M., Strom S.E.  1990, AJ 99, 869
\bibitem[1993]{kon:rud93}
K\"onigl A., Ruden S.P. 1993, in: Levy E.H., Lunine J.I. (eds.) 
Protostars and Planets III. University of Arizona Press, Tucson, p. 641
\bibitem[1987]{lada87}
Lada, C.J. 1987, in: Peimbert M., Jugaku J. (eds.) Star Forming Regions. IAU Symposium 115, p.1
\bibitem[1991]{Ladd91}
Ladd, E.F., Adams, F.C., Casey, S. et al. 1991, ApJ 382, 555
\bibitem[1980]{Landau}
Landau \& Lifshitz, 1987, Fluid Mechanics, 2nd edition, Pergamon, Oxford, p.358
\bibitem[1969]{Larson}
Larson, R.B. 1969, MNRAS, 145, 271
%\bibitem[1987]{LS87}
%Lizano, S., \& Shu, F.H. 1987, in; G.E. Morfill \& M. Scholer (eds.) 
%Physical Processes in Interstellar Clouds, p. 173
\bibitem[1989]{LS89}
Lizano, S., \& Shu, F.H. 1989, ApJ, 342, 834
%\bibitem[1989]{Loren89a}
%Loren, R.B. 1989a, ApJ, 338, 902
%\bibitem[1989]{Loren89b}
%Loren, R.B. 1989b, ApJ, 338, 925
%
\bibitem[1986]{LW86}
Loren, R.B., Wootten, A., 1986, ApJ, 306, 142
\bibitem[1990]{LWW90}
Loren, R.B., Wootten, A., Wilking, B.A. 1990, ApJ, 365, 229
%
\bibitem[1997]{Pudritz}
McLaughlin, D.E., \& Pudritz, R.E. 1997, ApJ, 476 
%
\bibitem[1992a]{Mezger92a}
Mezger, P.G., Sievers, A.W., Haslam, C.G.T., Kreysa, E., 
Lemke, R., Mauersberger, R., \& Wilson, T.L. 1992a, 
A\&A, 256, 631
%
\bibitem[1992b]{Mezger92b}
Mezger, P.G., Sievers, A.W., Zylka, R., Haslam, C.G.T., Kreysa, E., 
\& Lemke, R. 1992b, A\&A, 265, 743
%\bibitem[1994]{mizuno}
%Mizuno, A., Onishi, T., Hayashi, M., Ohashi, N., Sunada, K., Hasegawa, T., 
%Fukui, Y. 1994, Nature, 368, 719
\bibitem[1996]{motte96}
Motte, F., Andr\'e, P., Neri, R. 1996, in: Siebenmorgen, R., Kaufl, H.U.
(eds.) The Role of Dust in the Formation of Stars.  
ESO Astrophysics Symposia. Springer, Berlin, p.~47
\bibitem[1997]{motte97}
Motte, F., Andr\'e, P., Neri, R. 1997, A\&A, in preparation.
\bibitem[1989]{Mous89}
Mouschovias, T.M. 1989, in The Physics and Chemistry of Interstellar
Molecular Clouds, ed. G. Winnewisser \& J.T. Armstrong, 
Springer, Berlin, p.~297
\bibitem[1991]{Mous91}
Mouschovias, T.M. 1991, in The Physics of Star Formation and Early 
Stellar Evolution, Eds. Lada \& Kylafis, p. 449
\bibitem[1995]{Mous95}
Mouschovias, T.M. 1995, In: The Physics of the Interstellar Medium and Intergalactic
Medium, ed. A. Ferrara, C.F. Mc Kee, C. Heiles, \& P.R. Shapiro (San Francisco: ASP), 
Vol.~80, 184
%\bibitem[1983]{M83}
%Myers P. C. 1983, ApJ, 270, 105
\bibitem[1994]{M94}
Myers P. C. 1994, In: The Structure and Content of Molecular Clouds, ed. 
T.L. Wilson \& K.J. Johnston (Berlin: Springer),207
%\bibitem[1983]{MLB83}
%Myers P. C., Linke, R.A., \& Benson P. J., 1983, ApJ, 264, 517 -- MLB
%\bibitem[1992]{MF92}
%Myers P. C., \& Fuller, G. 1992, ApJ, 396, 631
%\bibitem[1988]{MG88}
%Myers P. C., \& Goodman, A.A. 1988, ApJ, 329, 392
%\bibitem[1991]{MG91}
%Myers P. C., \& Goodman, A.A. 1991, ApJ, 373, 509
%
\bibitem[1991]{MFGB91}
Myers P. C., Fuller, G., Goodman, A.A., \& Benson, P.J. 1991, ApJ, 376, 561
%
%\bibitem[1996]{Myers96}
%Myers P. C., Mardones, D., Tafalla, M., Williams, J.P., \& Wilner, D.J. 
%1996, ApJ, in press
%\bibitem[1995]{neri 95}
%Neri, R. et al. 1995, NIC users' manual (IRAM internal report)
\bibitem[1969]{Penston}
Penston, R.B. 1969, MNRAS, 144, 425
%\bibitem[1993]{Preibisch}
%Preibisch, Th., Ossenkopf, V., Yorke, H.W., Henning, Th. 1993, A\&A, 
%279, 577
%
\bibitem[1996]{Ryden}
Ryden, B.S. 1996, ApJ, 471, 822
\bibitem[1996]{sara96}
Saraceno P., Andr\'e P., Ceccarelli C., Griffin M., Molinari S. 1996, 
A\&A, 309, 827
\bibitem{b35}
Shu F. 1977, ApJ, 214, 488 -- Shu77
\bibitem{Shu95}
Shu F. 1995, Rev. Mex. Astr. Ap., Ser. de. Conf., 1, 375
\bibitem[1987]{shu87}
Shu, F.H., Adams, F.C., Lizano, S. 1987, ARA\&A 25,23
\bibitem[1993]{shu93}
Shu, F., Najita, J., Galli, D., Ostriker, E., \& Lizano S. 1993, 
in: Levy E.H., Lunine J.I. (eds.) 
Protostars and Planets III. University of Arizona Press, Tucson, p.~3
\bibitem[1994]{shu94}
Shu F., Najita J., Ostriker E. et al. 1994, ApJ 429, 781
\bibitem[1988]{stahler88}
Stahler S.W. 1988, ApJ 332, 804
\bibitem[1980]{stahler80}
Stahler, S.W., Shu, F.H., \& Taam, R.E. 1980, ApJ, 241, 637
\bibitem[1996]{Tomisaka96}
Tomisaka, K. 1996, PASJ, 48, L97
\bibitem[1988]{Tomisaka88}
Tomisaka, K., Ikeuchi, S., \& Nakamura, T. 1988, ApJ, 335, 239
\bibitem[1989]{Tomisaka89}
Tomisaka, K., Ikeuchi, S., \& Nakamura, T. 1989, ApJ, 341, 220
\bibitem[1996]{Troland96}
Troland, T.H., Crutcher, R.M., Goodman, A.A., Heiles, C., Kaz\`es, I., 
\& Myers, P.C. 1996, ApJ, 471, 302
\bibitem[1977]{Vrba} 
Vrba, F.J. 1977, AJ, 82, 198
%\bibitem[1976]{VSS} 
%Vrba, F.J., Strom, S.E., \& Strom, K. 1976, AJ, 81, 958
%\bibitem[1990]{WAL}
%Walker, C.K., Adams, F.C., \& Lada, C.J. 1990, ApJ, 349, 515
%\bibitem[1996]{WTJ}
%Ward-Thompson, D., Jessop, N. E., 1996, in: `Changing perceptions of the
%morphology, dust content and dust to gas ratios in galaxies', 
%ed: Block, D., Kluwer, Dordrecht, p.489
\bibitem[1994]{WSHA}
Ward-Thompson, D., Scott, P.F., Hills, R.E., \& Andr\'e, P. 1994, 
MNRAS, 268, 276 (WSHA)
\bibitem[1997]{WMA}
Ward-Thompson, D., Motte, F., \& Andr\'e, P. 1997, MNRAS, in preparation
\bibitem[1985]{WS}
Whitworth, A., \& Summers, D. 1985, MNRAS, 214, 1 -- WS85
\bibitem[1996]{Whitworth96}
Whitworth, A.P., Bhattal, A.S., Francis, N., \& Watkins, S.J. 1996, MNRAS, 
283, 1061
\bibitem[1989]{wly89}
Wilking B.A., Lada C.J., Young E.T. 1989, ApJ 340, 823
%\bibitem[1995]{Zhou}
%Zhou, S. 1995, ApJ, 442, 685
\bibitem[1984]{ZT}
Zinnecker, H., \& Tscharnuter, W.M. 1984, in: `Proceedings of the Workshop
on Star Formation', Ed. R.D. Wolstencroft, p. 83, Royal Observatory, Edinburgh

\end{thebibliography}
\end{document}